\documentclass[epj]{svjour}
\usepackage{amsmath}
\usepackage{graphicx}
\usepackage{amssymb}
\begin{document}
\title{\boldmath Non-Perturbative Contribution to the Thrust Distribution in
$e^+ e^-$ Annihilation}
\author{R.A.\ Davison\thanks{\emph{Address after 1 October 2008:} Rudolf Peierls Centre for Theoretical Physics, 1 Keble Road, Oxford OX1 3NP, UK}
\and B.R.\ Webber}
\institute{Cavendish Laboratory, J.J.\ Thomson Avenue, Cambridge CB3 0HE, UK}
\date{19 September 2008}

\abstract{We re-evaluate the non-perturbative contribution to the thrust
distribution in $e^+ e^-\to$ hadrons, in the light of the latest
experimental data and the recent NNLO perturbative calculation of this
quantity. By extending the calculation to NNLO+NLL accuracy, we
perform the most detailed study to date of the effects of
non-perturbative physics on this observable. In particular, we
investigate how well a model based on a low-scale QCD effective
coupling can account for such effects. We find that the difference
between the improved perturbative distribution and the experimental
data is consistent with a $1/Q$-dependent non-perturbative shift in
the distribution, as predicted by the effective coupling model. Best
fit values of
$\alpha_s\left(91.2\text{ GeV}\right)=0.1164^{+0.0028}_{-0.0026}$ and
$\alpha_0\left(2\text{ GeV}\right)=0.59\pm 0.03$ are obtained with
$\chi^2/\text{d.o.f.}=1.09$. This is consistent with NLO+NLL results
but the quality of fit is improved. The agreement in  $\alpha_0$
is non-trivial because a part of the $1/Q$-dependent contribution
(the infrared renormalon) is included in the NNLO perturbative
correction.
} 
\maketitle
\section{Introduction}
\label{sec:intro}
One of the most common and successful ways of testing QCD has been by
investigating the distribution of event shapes in $e^+e^-\rightarrow$
hadrons, which have been measured accurately over a range of
centre-of-mass energies ($14\text{ GeV}\le Q\le207\text{ GeV}$), and
provide a useful way of evaluating the strong coupling constant $\alpha_s$.

The main obstruction to obtaining an accurate value of $\alpha_s$ from
these distributions is not due to a lack of precise data but to
dominant errors in the theoretical calculation of the
distributions. In particular, there are non-perturbative effects that
cannot yet be calculated from first principles but cause
power-suppressed corrections that can be significant at experimentally
accessible energy scales. In the case of the thrust distribution
$d\sigma/dT$, previous work has shown that matching $\alpha_s$ with a
\textit{low-scale effective coupling} $\alpha_{\text{eff}}$ which
extrapolates below some infra-red matching scale $\mu_I$ results in a
$1/Q$-dependent \textit{shift} in the distribution that accounts well
for the discrepancy between the experimental and perturbative results
\cite{one}.

The presence of $1/Q$ corrections in event shapes is a generic
expectation based on the renormalon analysis of perturbation theory,
which implies an ambiguity of that order in the perturbative
predictions for these observables (see
\cite{Beneke:1998ui,Beneke:2000kc} for reviews).  The low-scale
effective coupling hypothesis \cite{Dokshitzer:1995qm} leads to
universality relations between the corrections to different
observables, valid to lowest order in the effective coupling,
and to a well-defined prescription for matching
the perturbative and non-perturbative contributions.

The calculation of Ref.~\cite{one} was performed to NLO+NLL accuracy,
i.e.\ terms up to ${\cal O}\left(\alpha_s^2\right)$ were retained exactly
while exponentiating logarithmically-enhanced terms of the form
$\alpha_s^n\ln^{n+1}(1-T)$ and $\alpha_s^n\ln^n(1-T)$ were summed to
all orders. In the present paper, the recent evaluation of the NNLO
term (i.e\ ${\cal O}\left(\alpha_s^3\right)$) in the fixed-order perturbation
series expansion of the thrust distribution \cite{two,three} is used
to refine the perturbative calculation of the distribution to NNLO+NLL
accuracy and thus to reduce the uncertainty present in the theoretical
prediction. A low-scale effective coupling is then introduced and
matched to NNLO.  This is again found to be a good method for dealing
with the non-perturbative shift. By comparing the NNLO+NLL+shift
results with the latest experimental distributions, values of
$\alpha_s$
and 
\begin{equation}
\alpha_0=\frac{1}{\mu_I}\int_0^{\mu_I}d\mu\text{ }\alpha_{\text{eff}}
\left(\mu\right)
\label{eqoneone}
\end{equation}
are obtained. These are consistent with those determined to
NLO+NLL accuracy. The agreement is non-trivial because a part of
the $1/Q$-dependent contribution -- the infrared renormalon -- is
included in the NNLO perturbative correction.

The organisation of the paper is as follows.  In Sect.~\ref{sec:thrust} we
briefly recall the relevant properties of the thrust distribution, the
fixed-order calculation and the resummation of large logarithms.
Sect.~\ref{sec:results} presents the predictions of perturbative
NNLO+NLL matching and the power dependence of the discrepancy with
experimental data. The matching to the low-scale effective coupling
and comparisons with data are performed in Sect.~\ref{sec:NP}, and our
conclusions are presented in Sect.~\ref{sec:conc}.

\section{Perturbative calculation of the thrust distribution}
\label{sec:thrust}
We recall that the thrust $T$ is a measure of the distribution of momenta of the final state hadrons:
\begin{equation}
T=\max_{\overrightarrow{n}}\left(\frac{\sum_{i=1}^N\left|\overrightarrow{p_i}.\overrightarrow{n}\right|}{\sum_{i=1}^N\left|\overrightarrow{p_i}\right|}\right),
\label{eqthreeone}
\end{equation}
where $\overrightarrow{n}$ is a unit vector and we sum over the
3-momentum of each final-state hadron in the centre-of-mass
frame. Theoretical calculations of thrust are performed by summing
over the individual final state partons, as the hadronisation process
is still not well understood. $T$ can vary between the limits $T=1$
for back-to-back jets and $T=\frac{1}{2}$ for a uniform angular
distribution of
hadrons.

For comparison with experiments, it is the thrust distribution
\begin{equation}
\frac{1}{\sigma}\frac{d\sigma}{dT},
\label{eqthreetwo}
\end{equation}
which is relevant, where $\sigma$ is the total cross-section for
$\text{e}^+\text{e}^-\rightarrow$ hadrons. In calculations it is more
convenient to use the event shape variable
\begin{equation}
t\equiv1-T,
\label{eqthreethree}
\end{equation}
which has the two-jet limit $t=0$. The distribution away from this
limit therefore depends directly upon the production of extra
final-state partons at QCD vertices, and hence is ideal for testing
QCD and evaluating $\alpha_s$.  The normalised thrust cross section
is then defined as
\begin{equation}
R\left(t\right)=\int_0^tdt\frac{1}{\sigma}\frac{d\sigma}{dt}=\int_{1-t}^1dT\frac{1}{\sigma}\frac{d\sigma}{dT}.
\label{eqthreefour}
\end{equation}

\subsection{Fixed-order calculations}
\label{sec:pert}
The perturbative expansion of the normalised thrust cross section has
the
general form
\begin{equation}
R(t) = 1+\bar{\alpha}_s R_1(t)+\bar{\alpha}_s^2 R_2(t)+\bar{\alpha}_s^3 R_3(t)+\ldots,
\label{eqthreefive}
\end{equation}
where $R_1\left(t\right)$ is the leading order (LO) coefficient,
$R_2\left(t\right)$ is the next-to-leading order (NLO) coefficient,
$R_3\left(t\right)$ is the next-to-next-to-leading order (NNLO)
coefficient etc. and $\bar{\alpha}_s\equiv\alpha_s/2\pi$.
Solving the renormalisation group equation for the running coupling to NNLO gives
\begin{eqnarray}
\alpha_s\left(\mu_{R}\right)&=&\frac{2\pi}{\beta_0L}\Biggl(1-\frac{\beta_1\ln
    L}{\beta_0^2L}+\frac{1}{\beta_0^2L^2}\nonumber\\
&&\left[\frac{\beta_1^2}{\beta_0^2}\left(\ln^2L-\ln L-1\right)+\frac{\beta_2}{\beta_0}\right]\Biggr),
\label{eqtwosix}
\end{eqnarray}
where $\mu_{R}$ is some chosen renormalisation scale (we take $\mu_R=Q$ except
where stated otherwise),
\begin{equation}
\begin{aligned}
\beta_0&=\frac{11N-2N_F}{6}\;\qquad \beta_1=\frac{17N^2-5NN_F-3C_FN_F}{6},\\
\beta_2&=\frac{1}{432}(2857N^3+54C_F^2N_F-615NC_FN_F\\
& -1415N^2N_F+66C_FN_F^2+79NN_F^2)\;,
\end{aligned}
\label{eqtwoseven}
\end{equation}
with $C_F=(N^2-1)/2N$ for an SU($N$) gauge theory with $N_F$ active
flavours ($N=3$ for QCD and $N_F=5$ at all energies considered here)
and $L=\ln(\mu_{R}^2/\Lambda_{\overline{MS}}^{\left(5\right)\text{
  } 2})$, $\Lambda_{\overline{MS}}^{\left(5\right)}$ being the
5-flavour QCD scale in the modified minimal subtraction
renormalisation scheme.

\begin{figure}
\begin{center}
\includegraphics[scale=0.51]{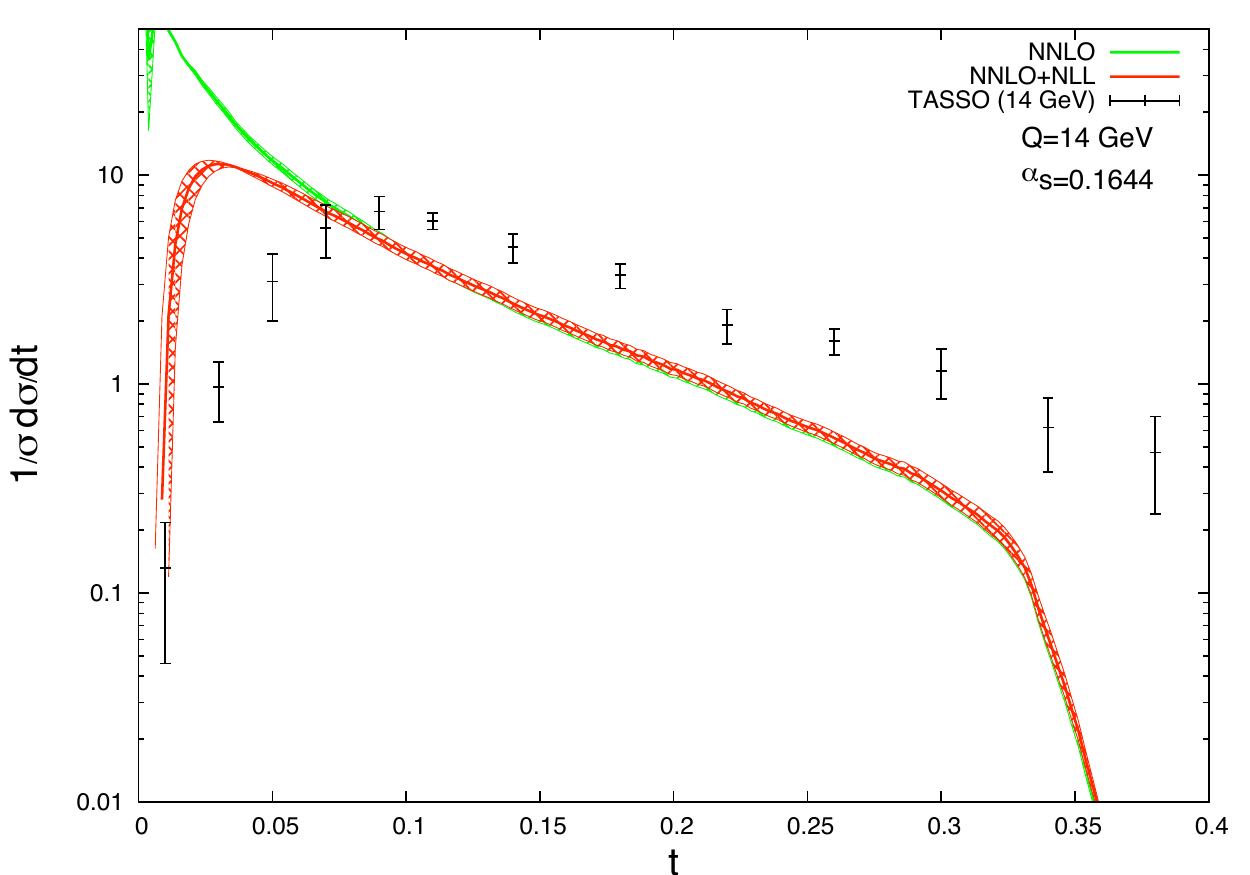}
\includegraphics[scale=0.51]{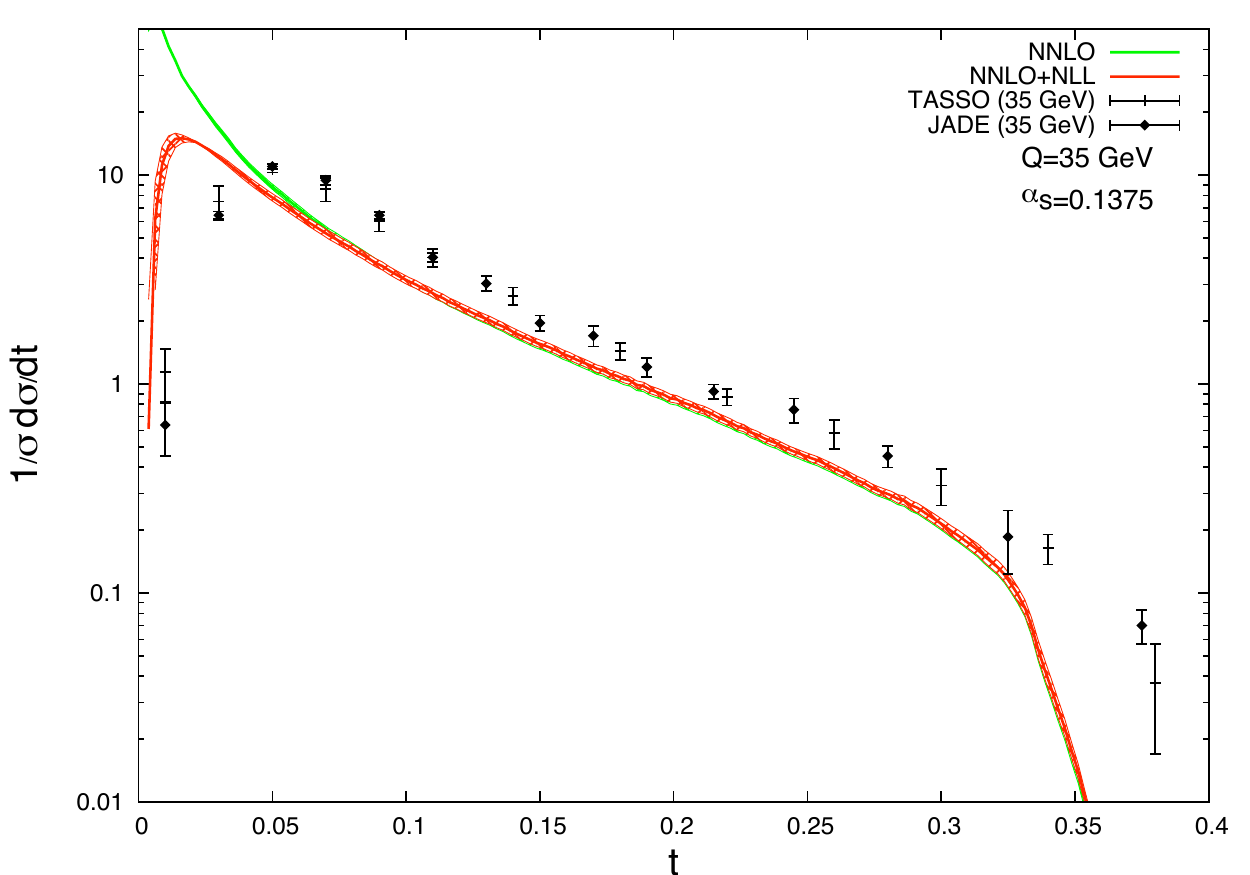}
\includegraphics[scale=0.51]{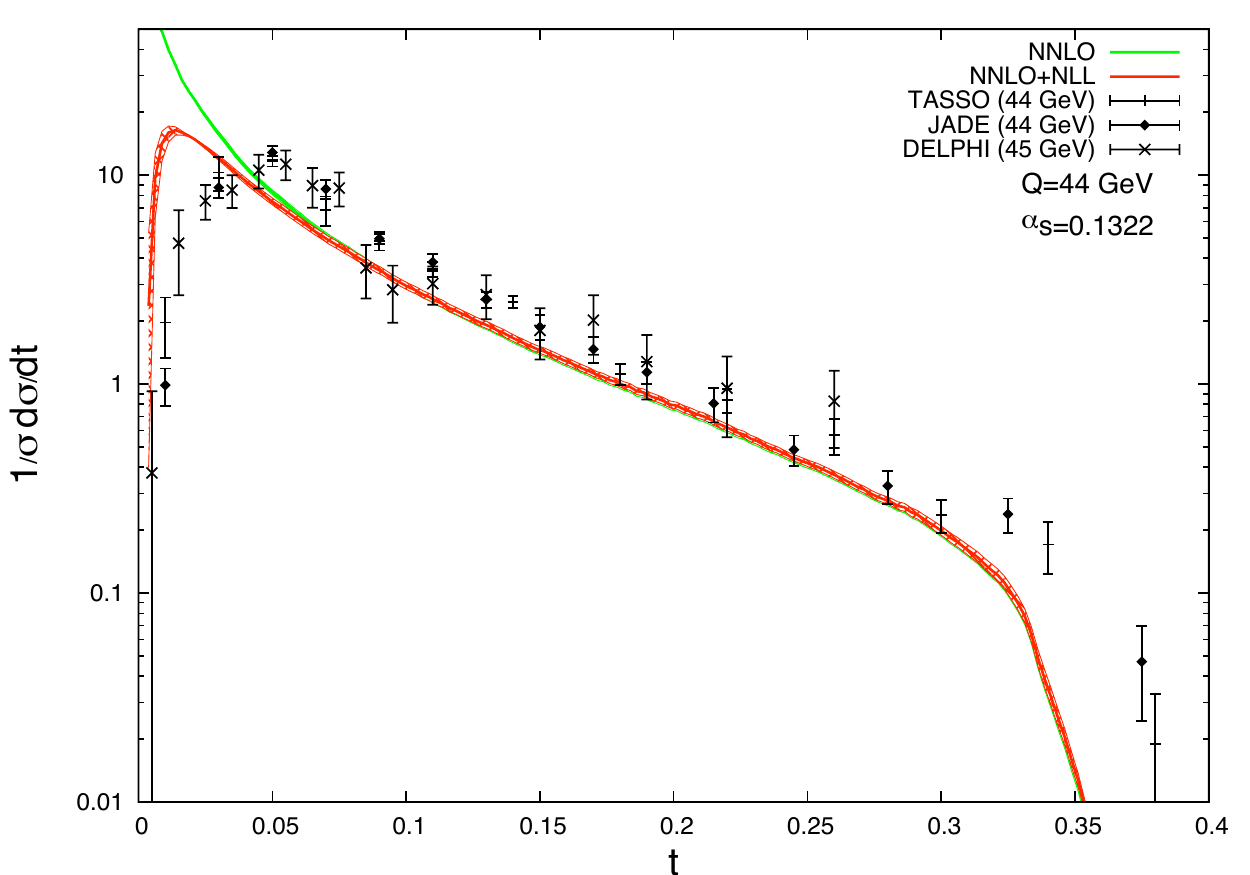}
\includegraphics[scale=0.51]{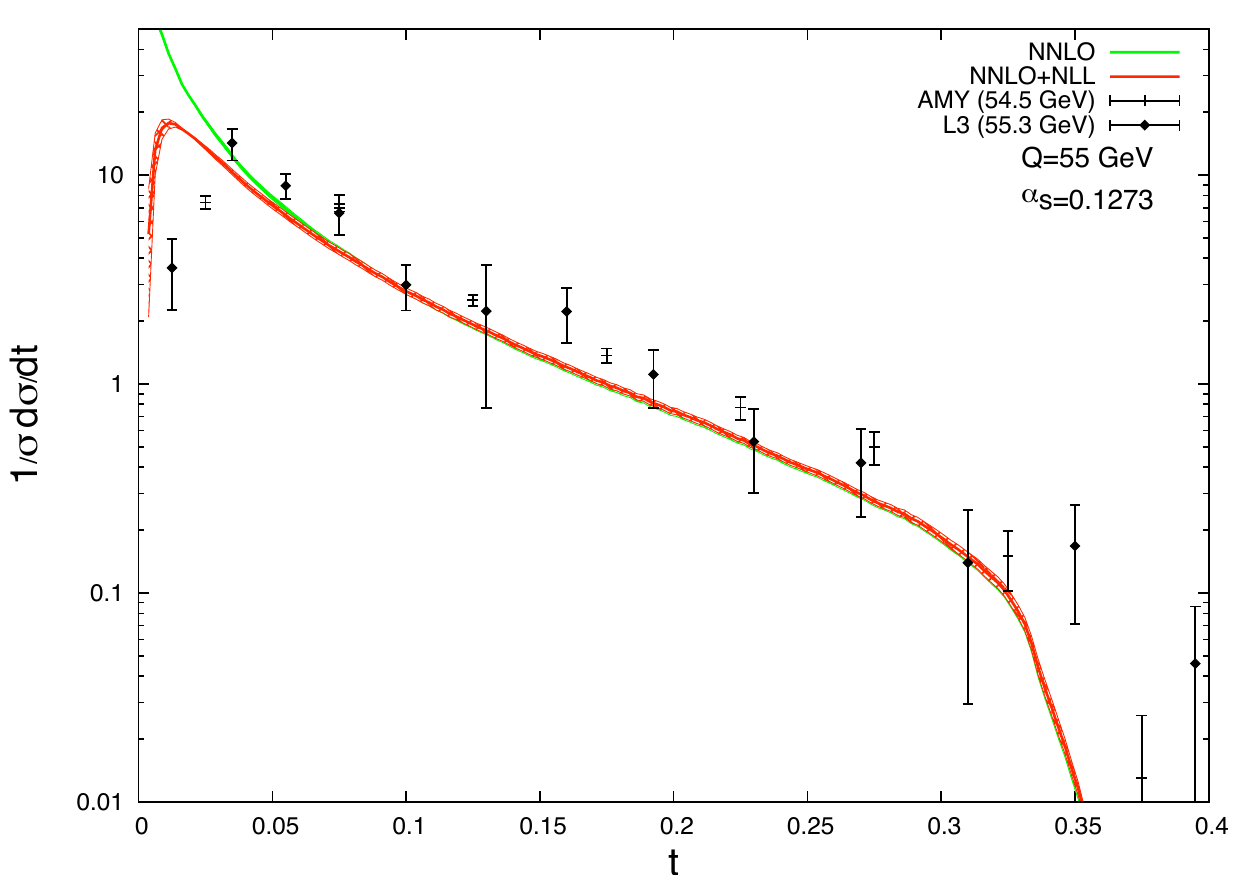}
\includegraphics[scale=0.51]{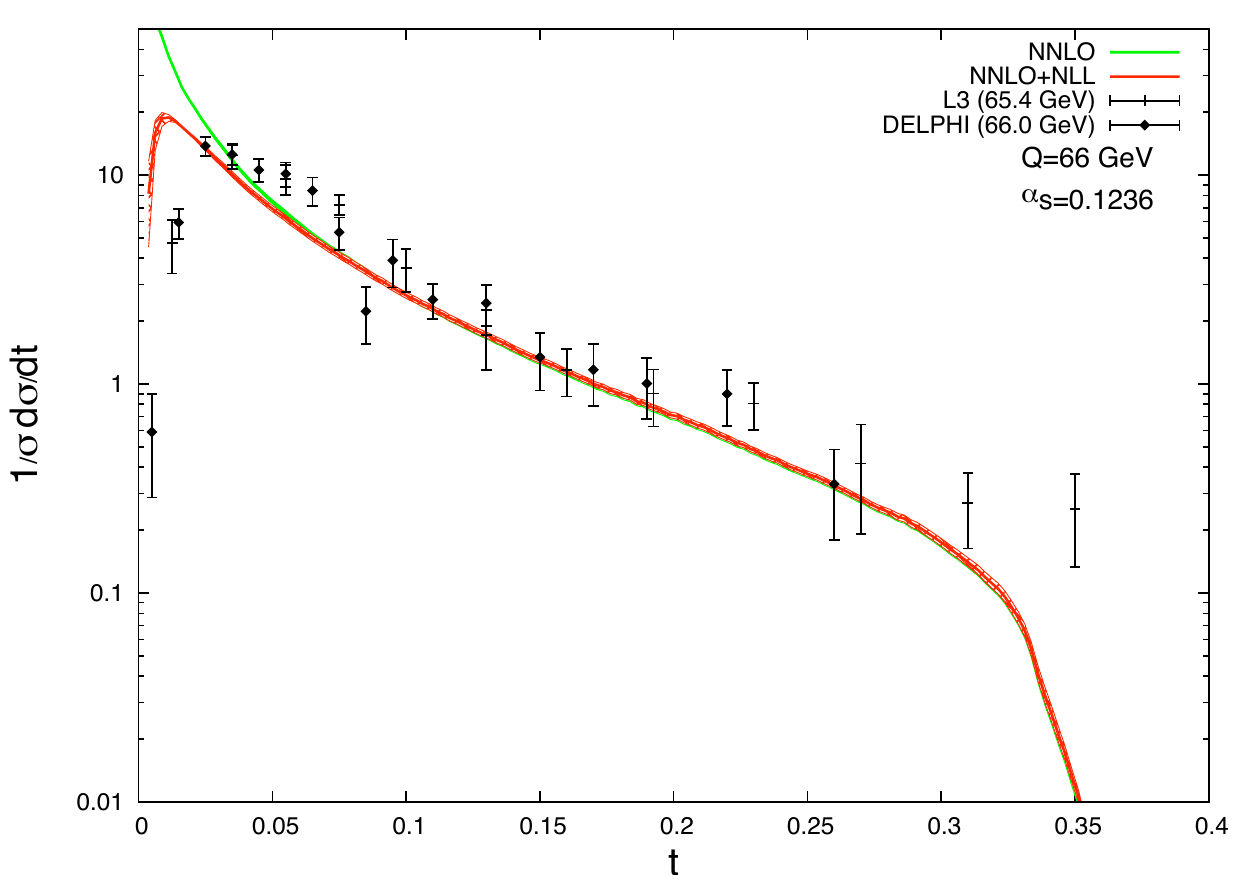}
\caption{Fixed-order (NNLO), resummed (NNLO+NLL)
and experimental thrust distributions: $Q=14-66$ GeV.\label{fig:unshift1}}
\end{center}
\end{figure}

\begin{figure}
\begin{center}
\includegraphics[scale=0.51]{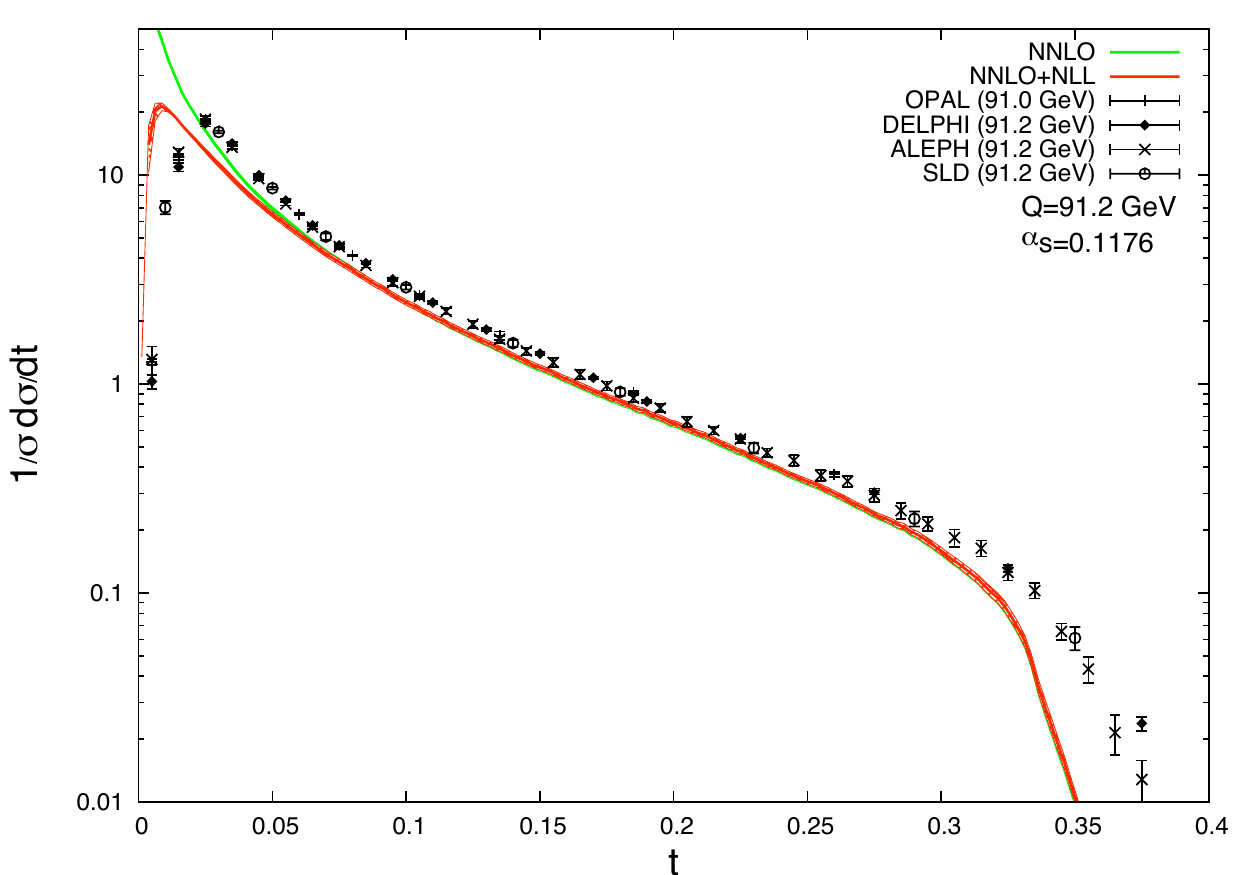}
\includegraphics[scale=0.51]{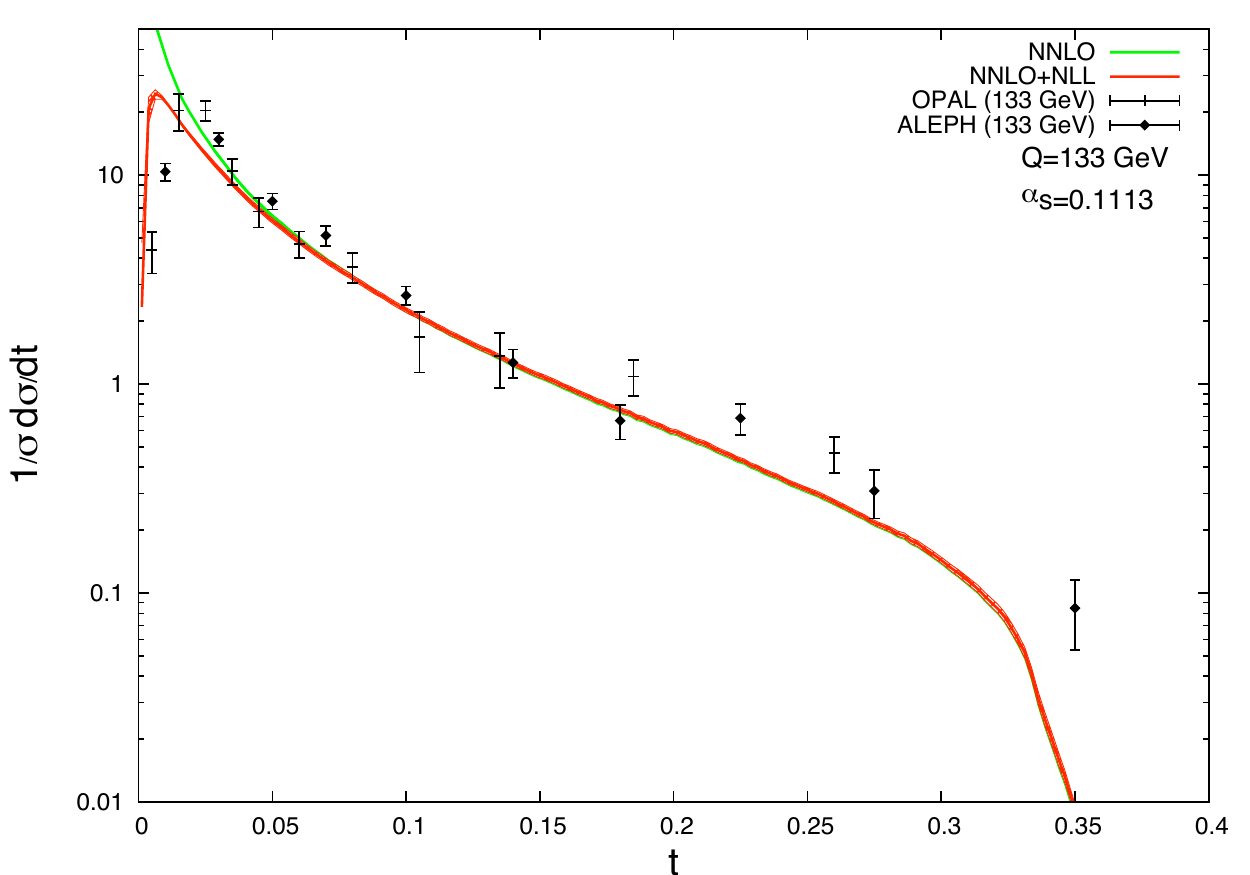}
\includegraphics[scale=0.51]{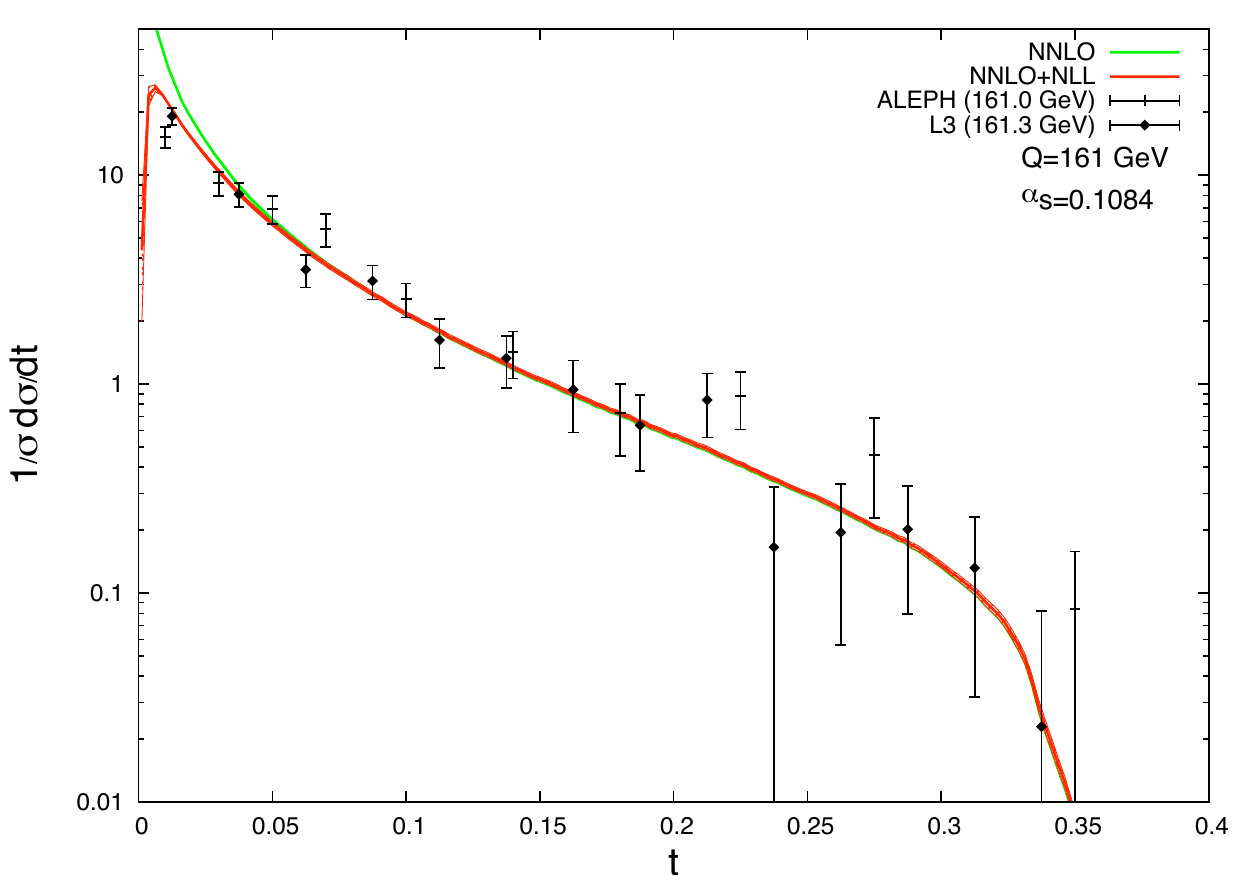}
\includegraphics[scale=0.51]{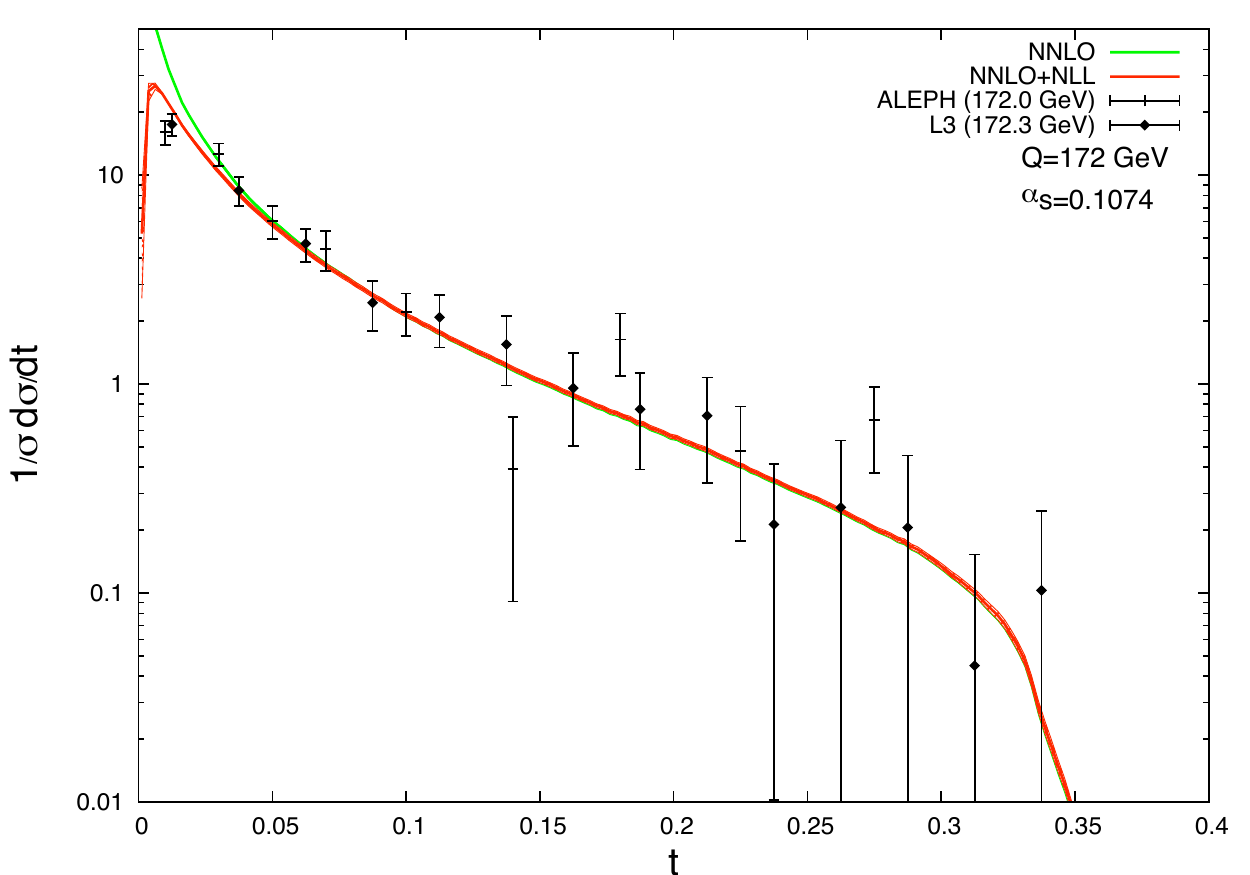}
\includegraphics[scale=0.51]{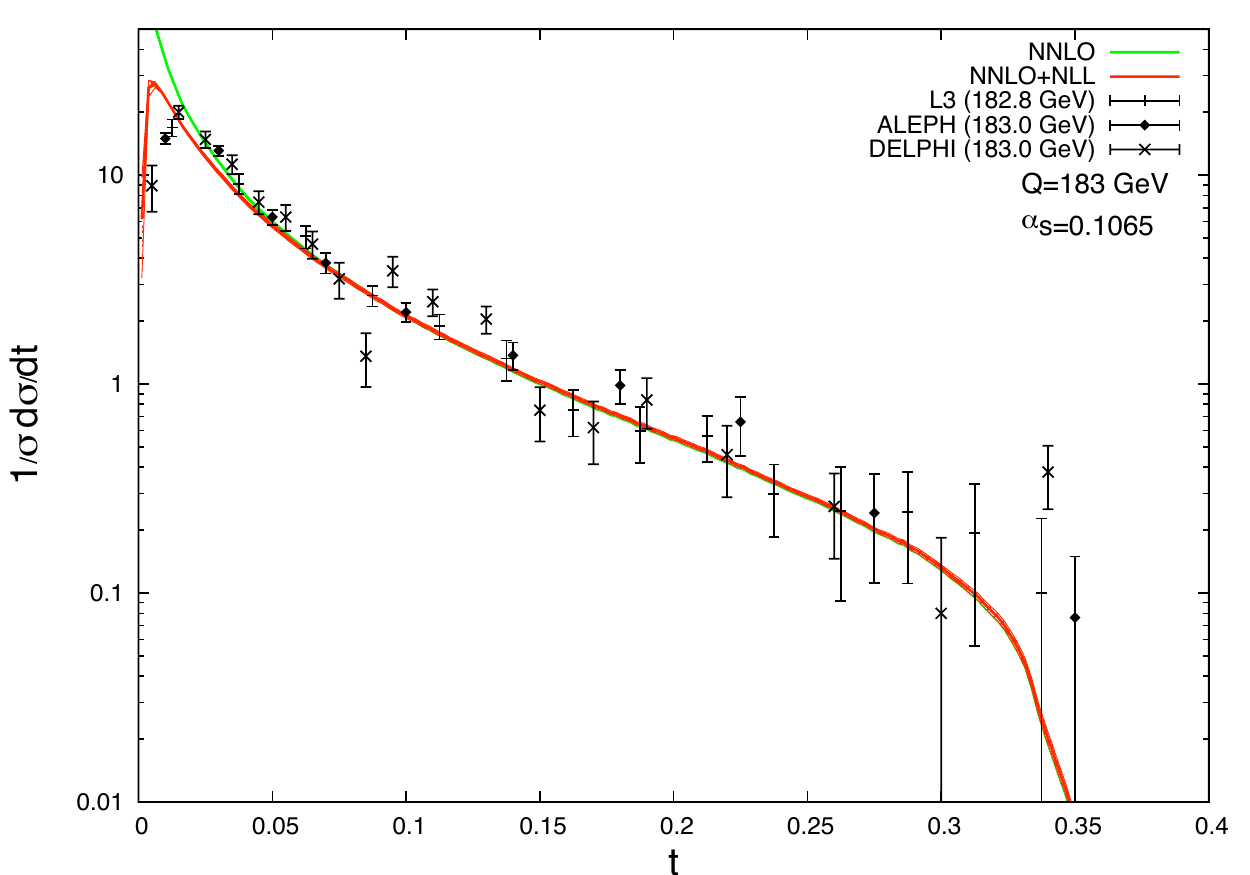}
\caption{Fixed-order (NNLO), resummed (NNLO+NLL)
and experimental thrust distributions: $Q=91-183$ GeV.\label{fig:unshift2}}
\end{center}
\end{figure}

\begin{figure}
\begin{center}
\includegraphics[scale=0.51]{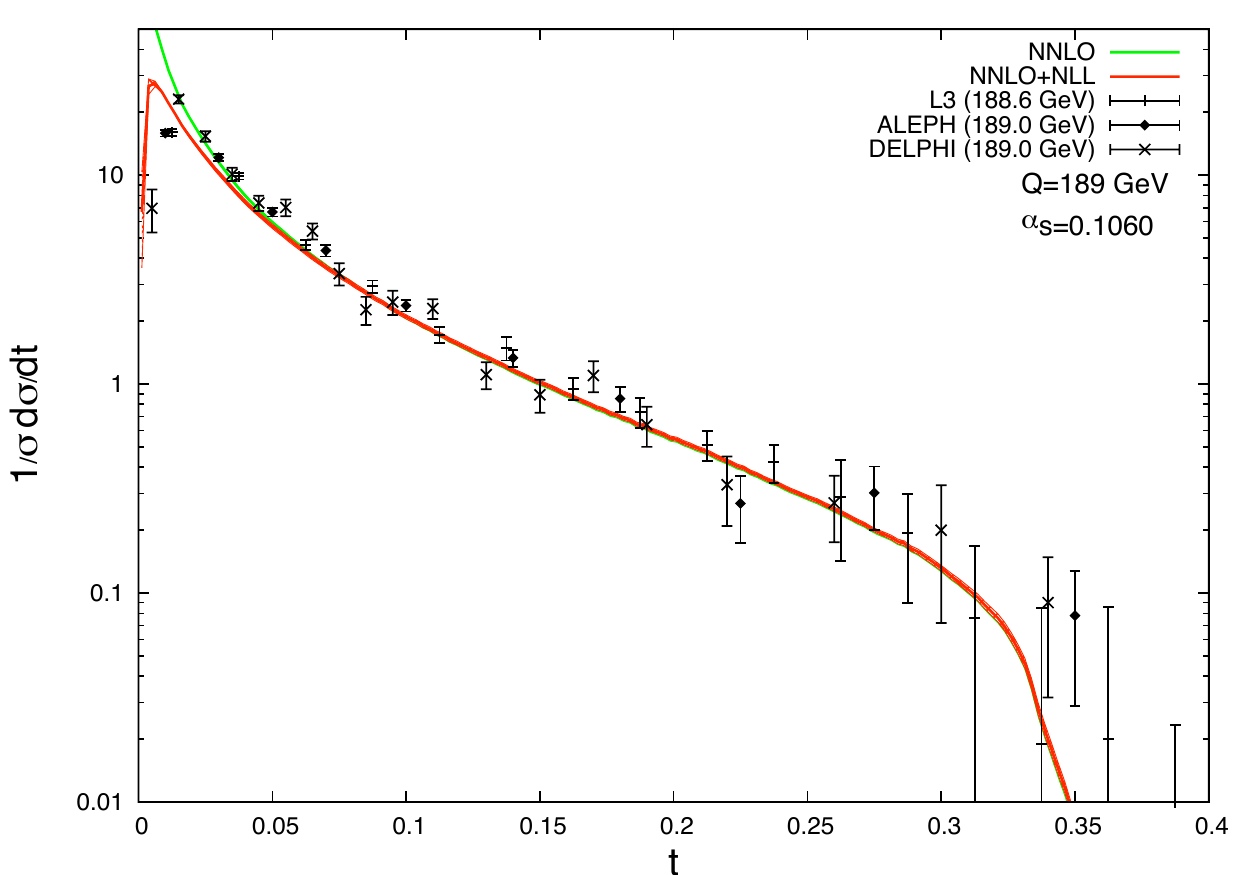}
\includegraphics[scale=0.51]{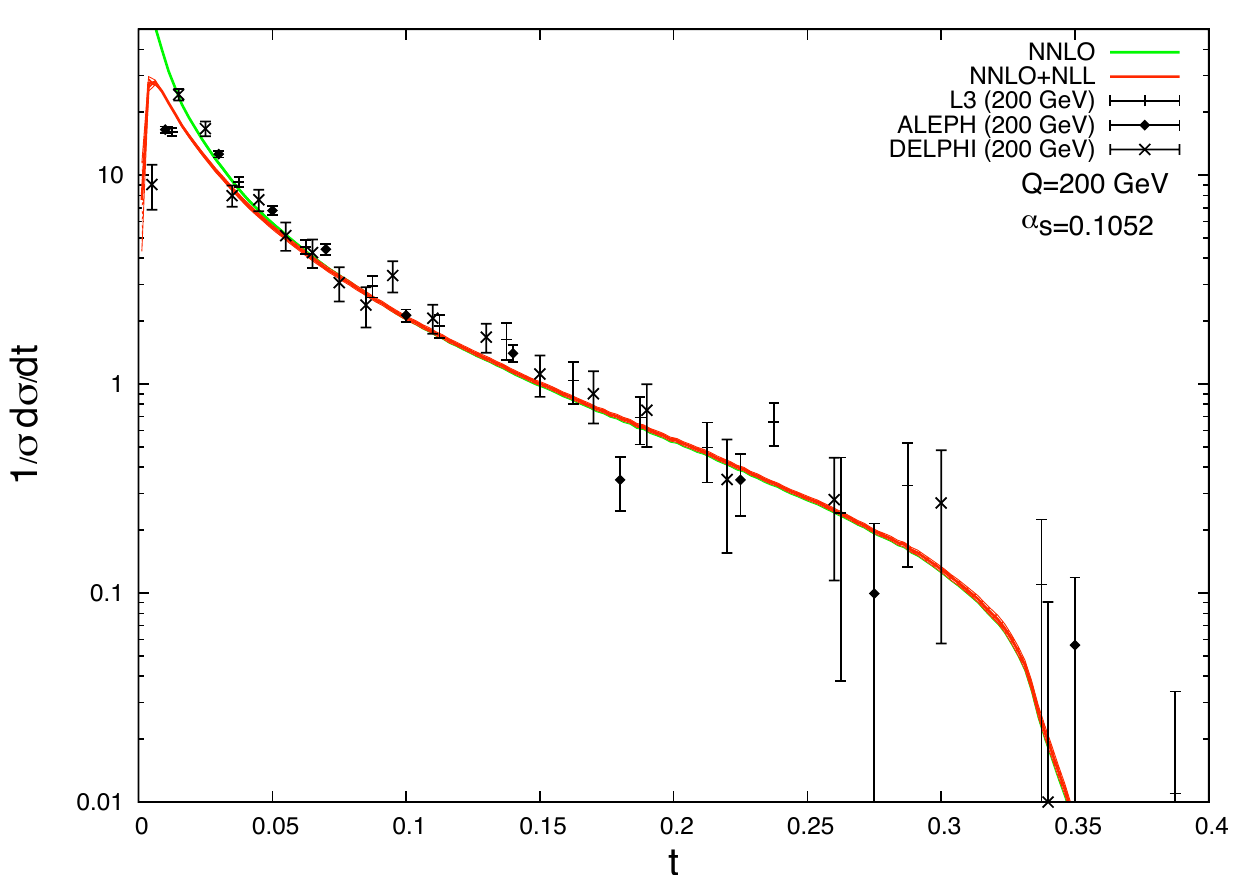}
\includegraphics[scale=0.51]{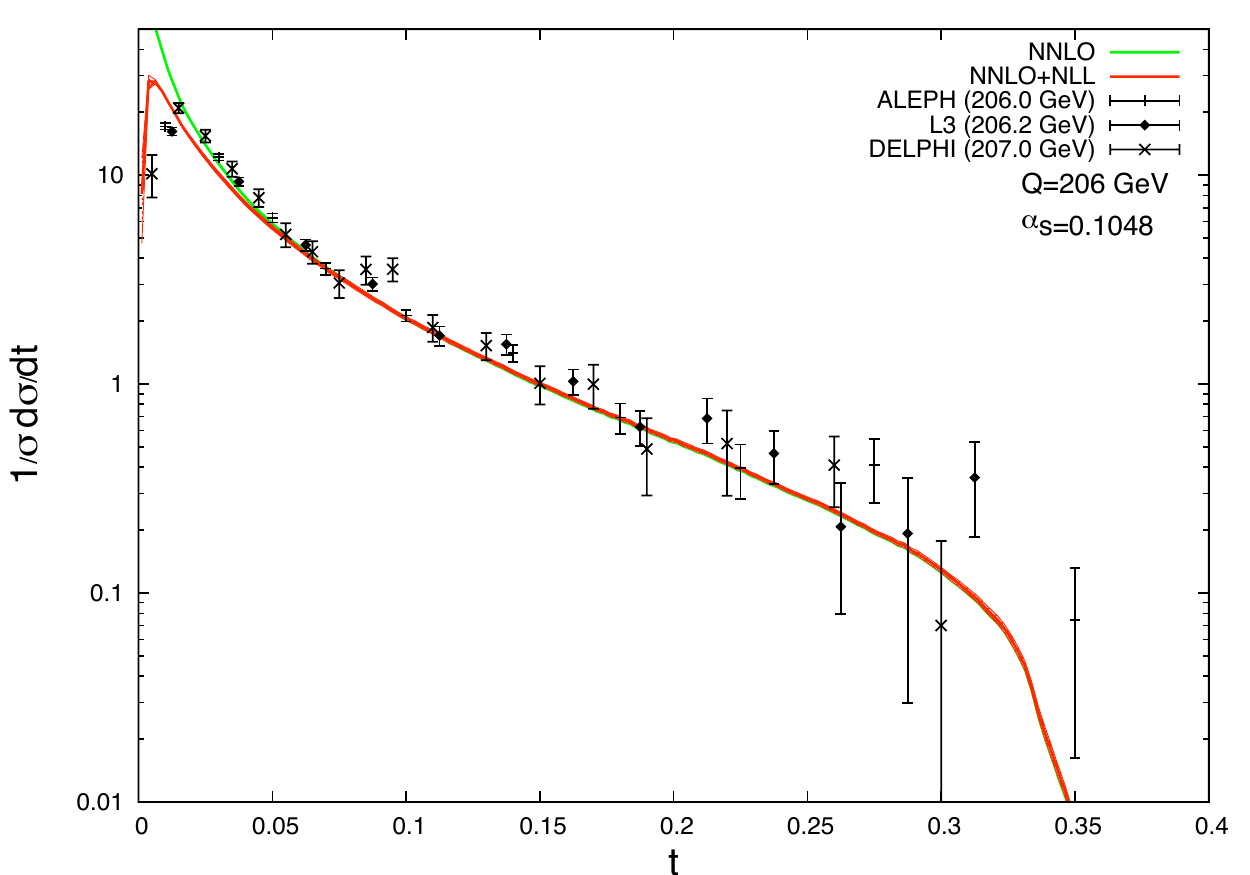}
\caption{Fixed-order (NNLO), resummed (NNLO+NLL)
and experimental thrust distributions: $Q=189-207$ GeV.\label{fig:unshift3}}
\end{center}
\end{figure}

A numerical Monte Carlo program,
\texttt{EERAD3}~\cite{GehrmannDeRidder:2007jk}, has recently been developed
which computes the process $e^+e^-\rightarrow$ jets to NNLO
in $\alpha_s$ via the decay of a virtual neutral gauge boson ($\gamma$
or $Z^0$) to between three and five partons
\cite{two,three}.\footnote{A recent calculation \cite{Weinzierl:2008iv} finds 
some discrepancies with Refs.~\cite{two,three}, but these are not
significant in the kinematic regions that we consider.}
The \texttt{EERAD3} predictions for the thrust distribution at a variety of
centre-of-mass energies $Q$ spanning the range 14 GeV to 206 GeV are shown
by the green/lighter curves in Figs.~\ref{fig:unshift1}-\ref{fig:unshift3}.
The values of $\alpha_s\left(Q\right)$ were calculated using 
$\Lambda_{\overline{MS}}^{\left(5\right)}=0.204$ GeV, corresponding
to the world average
$\alpha_s\left(91.2\;\text{GeV}\right)=0.1176$ \cite{Amsler:2008zz}.


\subsection{Resummation of large logarithms}
\label{sec:resum}
The enhancement of the distribution at low $t$ due to soft or
collinear gluon emission (as seen in Figs.~\ref{fig:unshift1}-
\ref{fig:unshift3}) is present at all orders in perturbation
theory: the dominant term at $n$th order is typically of the form
\begin{equation}
\label{eqfourone}
\frac{1}{\sigma}\frac{d\sigma}{dt}\sim\alpha_s^n\frac{1}{t}\ln^{2n-1}\left(\frac{1}{t}\right).
\end{equation}

Thus we see that at low $t$ the condition $\alpha_s\ll1$ is not
sufficient for a fixed-order prediction in perturbation theory to be
accurate. Instead, we require $\alpha_sL^2\ll1$, where
$L\equiv\ln(1/t)$. To obtain accurate predictions in the two-jet limit
$t\rightarrow 0$, we must therefore take account of these enhanced
terms at all orders in perturbation theory by resumming them. 

Resummation of large logarithms is possible for event shape variables
$y$ that \textit{exponentiate} \cite{four}, i.e.\ their corresponding
normalised cross section can be written in the form
\begin{equation}
R\left(y\right)=C\left(\alpha_s\right)\Sigma\left(y,\alpha_s\right)+D\left(y,\alpha_s\right),
\label{eqfourtwo}
\end{equation}
where
\begin{equation}
\begin{aligned}
C\left(\alpha_s\right)&=1+\sum_{n=1}^{\infty}C_n\bar{\alpha}_s^n,
\\
\ln\Sigma\left(y,\alpha_s\right)&=\sum_{n=1}^{\infty}\sum_{m=1}^{n+1}G_{nm}\bar{\alpha}_s^nL^m
\\
&=Lg_1\left(\alpha_sL\right)+g_2\left(\alpha_sL\right)+\alpha_sg_3\left(\alpha_sL\right)+\ldots,
\end{aligned}
\label{eqfourthree}
\end{equation}
$L=\ln(1/y)$ and $D\left(y,\alpha_s\right)$ is a remainder function
that vanishes order-by-order in perturbation theory in the two-jet limit
$y\rightarrow0$. The functions $g_i\left(\alpha_sL\right)$ are power
series in $\alpha_sL$ (with no leading constant term) and hence
$Lg_1\left(\alpha_sL\right)$ sums all leading logarithms
$\alpha_s^{n}L^{n+1}$, $g_2\left(\alpha_sL\right)$ sums all
next-to-leading logarithms (NLL) $\alpha_s^nL^n$ and the
subdominant logarithmic terms $\alpha_s^nL^m$ with $0<m<n$
are contained in the $g_3, g_4,\ldots$ terms. The functions $g_i$ thus
resum the logarithmic contributions at all orders in
perturbation theory, and knowledge of their form allows us to make
accurate perturbative predictions in the range $\alpha_sL\lesssim 1$
 -- a significant improvement on the fixed-order range $\alpha_sL^2\ll1$.

For thrust, the first two functions can be determined analytically by
using the \textit{coherent branching}
formalism~\cite{Dokshitzer:1982xr,Bassetto:1984ik}, which uses
consecutive branchings from an initial quark-antiquark state to
produce multi-parton final states to NLL accuracy. The results of this
calculation depend upon the \textit{jet mass distribution}
$J\left(Q^2,k^2\right)$ -- the probability of producing a final state
jet with invariant mass $k^2$ from a parent parton produced in a hard
process at scale $Q^2$ -- and its Laplace transform
$\tilde{J}_{\nu}\left(Q^2\right)$. To the required accuracy,
the thrust distribution is
\begin{equation}
\frac{1}{\sigma}\frac{d\sigma}{dt}=\frac{Q^2}{2\pi i}\int_Cd\nu e^{t\nu Q^2}\left[\tilde{J}_{\nu}^{\mu}\left(Q^2\right)\right]^2,
\label{eqfourfour}
\end{equation}
where the contour $C$ runs parallel to the imaginary axis on the right
of all singularities of the integrand, 
\begin{eqnarray}
\ln\tilde{J}_{\nu}^{\mu}\left(Q^2\right)&=&\int_0^1\frac{du}{u}\left(e^{-u\nu
    Q^2}-1\right)\Biggl[\int_{u^2Q^2}^{uQ^2}\frac{d\mu^2}{\mu^2}C_F\frac{\alpha_s\left(\mu\right)}{\pi}\nonumber\\
&&\left(1-K\frac{\alpha_s\left(\mu\right)}{2\pi}\right)^{-1}+\ldots\Biggr],
\label{eqfourfive}
\end{eqnarray}
and\footnote{By writing the $K$ dependence in the form shown in
(\ref{eqfourfive}), we change from the $\overline{\mbox{MS}}$
renormalisation scheme to the so-called bremsstrahlung scheme~\cite{Catani:1990rr}.}
\begin{equation}
\begin{aligned}
K=N\left(\frac{67}{18}-\frac{\pi^2}{6}\right)-\frac{5}{9}N_F\;.
\end{aligned}
\label{eqfoursix}
\end{equation}

This expression demonstrates explicitly that the divergence of
$\alpha_s\left(\mu\right)$ at low $\mu$ will affect the perturbative
thrust distribution -- such effects are related to the renormalon
mentioned earlier. To NLL accuracy, however, we can neglect the low
$\mu$ region (although we will return to it in Sect.~\ref{sec:NP}) to
give the thrust resummation functions~\cite{four}
\begin{equation}
\begin{aligned}
&g_1\left(\alpha_sL\right)=2f_1\left(\beta_0\bar{\alpha}_sL\right),
\\
&g_2\left(\alpha_sL\right)=2f_2\left(\beta_0\bar{\alpha}_sL\right)-\ln\Gamma[1-2f_1\left(\beta_0\bar{\alpha}_sL\right)\\
&\;\;\;\;-2\beta_0\bar{\alpha}_sLf_1'\left(\beta_0\bar{\alpha}_sL\right)],
\end{aligned}
\label{eqfourseven}
\end{equation}
where
\begin{equation}
\begin{aligned}
f_1\left(x\right)=&-\frac{C_F}{\beta_0x}[\left(1-2x\right)\ln\left(1-2x\right)\\
&-2\left(1-x\right)\ln\left(1-x\right)],
\\
f_2\left(x\right)=&-\frac{C_FK}{\beta_0^2}\left[2\ln\left(1-x\right)-\ln\left(1-2x\right)\right]
\\
&-\frac{3C_F}{2\beta_0}\ln\left(1-x\right)-\frac{2C_F\gamma_E}{\beta_0}[\ln\left(1-x\right)\\
&-\ln\left(1-2x\right)]-\frac{C_F\beta_1}{\beta_0^3}\bigl[\ln\left(1-2x\right)\\
&-2\ln\left(1-x\right)+\frac{1}{2}\ln^2\left(1-2x\right)-\ln^2\left(1-x\right)\bigr],
\end{aligned}
\label{eqfoureight}
\end{equation}
with $\Gamma$ the Euler $\Gamma$-function, $\gamma_E$ the Euler constant, and $C_F$, $K$ and $\beta_n$ the constants previously defined.

By combining these with the fixed-order calculation, we can obtain a
new estimate of the normalised cross section to NLL accuracy. This
should particularly improve the fixed-order estimate in the two-jet
region, where $L$ becomes large. Naively we would
simply calculate $R\left(t\right)$ as defined in
Eq.~(\ref{eqfourtwo}), but it turns out to be considerably simpler
to consider $\ln R\left(t\right)$, as we recall next.

\subsection{Log-R matching}
\label{sec:match}
In the \textit{log-R matching scheme}, we rewrite the exponentiation formula as
\begin{equation}
\ln R\left(t\right)=F\left(\alpha_s\right)+\ln\Sigma\left(t,\alpha_s\right)+H\left(t,\alpha_s\right),
\label{eqfournine}
\end{equation}
where $F\left(\alpha_s\right)$ is a power series in $\alpha_s$ and $H\left(t,\alpha_s\right)$ denotes the remainder function which vanishes as $t\rightarrow0$.

For a fixed-order perturbative calculation of $R\left(t\right)$ to
order $M$, we can write Eq.~(\ref{eqthreefive}) as

\begin{equation}
\begin{aligned}
\ln
R\left(t\right)&=\ln\left(1+\sum_{n=1}^M\bar{\alpha}_s^nR_n\left(t\right)\right)\\
&=\sum_{n=1}^M\bar{\alpha}_s^nR_n\left(t\right)-\frac{1}{2}\left(\sum_{n=1}^M\bar{\alpha}_s^nR_n\left(t\right)\right)^2\\
&+\frac{1}{3}\left(\sum_{n=1}^M\bar{\alpha}_s^nR_n\left(t\right)\right)^3-\ldots.
\end{aligned}
\label{eqfourten}
\end{equation}
The matched estimate is obtained by combining the $M$th order perturbative result with the resummed contributions and subtracting the terms of order $\leq M$ in $\ln\Sigma$ (as these are already accounted for in the fixed-order terms). Thus for a fixed-order calculation to order $\alpha_s^3$, the matched estimate after resumming large logarithms to NLL accuracy is
\begin{equation}
\begin{aligned}
\ln R\left(t\right)&=Lg_1\left(\alpha_sL\right)+g_2\left(\alpha_sL\right)\\
&+\bar{\alpha}_s\left(R_1\left(t\right)-G_{11}L-G_{12}L^2\right)\\
&+\bar{\alpha}_s^2\left(R_2\left(t\right)-\frac{1}{2}\left[R_1\left(t\right)\right]^2-G_{22}L^2-G_{23}L^3\right)\\
&+\bar{\alpha}_s^3\biggl(R_3\left(t\right)-R_1\left(t\right)R_2\left(t\right)+\frac{1}{3}\left[R_1\left(t\right)\right]^3\\
&-G_{33}L^3-G_{34}L^4\biggr).
\end{aligned}
\label{eqfoureleven}
\end{equation}

The coefficients $G_{nm}$ can be extracted by expanding the functions $g_1\left(\alpha_sL\right)$ and $g_2\left(\alpha_sL\right)$ as power series in $\alpha_sL$ and comparing them with the definition (\ref{eqfourthree}) of $G_{nm}$:
\begin{equation}
\begin{aligned}
G_{11}&=3C_F,
\\
G_{12}&=-2C_F,
\\
G_{22}&=-\frac{C_F}{36}\left[48\pi^2C_F+\left(169-12\pi^2\right)N-22N_F\right],
\\
G_{23}&=-\frac{C_F}{3}\left(11N-2N_F\right),
\\
G_{33}&=\frac{C_F}{108}\bigl[2304\zeta\left(3\right)C_F^2-792\pi^2NC_F\\
&-\left(3197-132\pi^2\right)N^2+\left(108+144\pi^2\right)C_FN_F\\
&+\left(1024-24\pi^2\right)NN_F-68N_F^2\bigr],
\\
G_{34}&=-\frac{7}{108}C_F\left(11N-2N_F\right)^2,
\end{aligned}
\label{eqfourtwelve}
\end{equation}
where $\zeta\left(3\right)=1.202057\ldots$.

There are two reasons why it is simpler to use this log-$R$ matching
scheme rather than $R$ matching (i.e. evaluating Eq.~(\ref{eqfourtwo})
explicitly to NLL precision). Firstly, we do not have to be concerned
with the $C\left(\alpha_s\right)$ and $D\left(t,\alpha_s\right)$ terms
in (\ref{eqfourtwo}), for which we do not have analytic expressions but
which contribute to the fixed-order calculation -- these are contained
in $R_1\left(t\right)$, $R_2\left(t\right)$, etc. Secondly, it is
easier to impose physical boundary conditions on the normalised cross
section, namely
\begin{equation}
R\left(t=t_{max}\right)=1,
\label{eqfourthirteen}
\end{equation}
by definition of the normalised cross section, and
\begin{equation}
\frac{dR}{dt}\left(t=t_{max}\right)=0,
\label{eqfourfourteen}
\end{equation}
as there is an upper kinematic limit $t_{\text{max}}$ on the thrust
for a given number of final-state partons. Although the resummed
logarithmic terms are small at high $t$, $dR/dt$ is also small and
so these terms can cause relatively large unphysical effects if we
do not impose these conditions. 

The above constraints are automatically obeyed by the fixed-order terms $R_n\left(t\right)$ but \textit{not} by the resummed terms, as we have neglected the subdominant logarithms $g_3\left(\alpha_sL\right)$, $g_4\left(\alpha_sL\right)$ etc. To satisfy these constraints, we therefore require
\begin{equation}
\begin{aligned}
Q\left(t\right)&=Lg_1\left(\alpha_sL\right)+g_2\left(\alpha_sL\right)-\bar{\alpha}_s\left(G_{11}L+G_{12}L^2\right)\\&-\bar{\alpha}_s^2\left(G_{22}L^2+G_{23}L^3\right)-\bar{\alpha}_s^3\left(G_{33}L^3+G_{34}L^4\right)
\end{aligned}
\label{eqfourfifteen}
\end{equation}
and its first derivative to vanish at $t=t_{max}$. $Q\left(t\right)$ corresponds to the resummed logarithmic terms of order $L^4$ and higher and hence at small $L$,
\begin{equation}
t\frac{dQ}{dt}=aL^3+bL^4+cL^5+\ldots.
\label{eqfoursixteen}
\end{equation}
By making the replacement
\begin{equation}
L\rightarrow\tilde{L}=\ln\left(1+\frac{1}{t}-\frac{1}{t_{max}}\right),
\label{eqfourseventeen}
\end{equation}
the boundary conditions are satisfied as $\tilde{L}\left(t_{max}\right)=0$. This does introduce corrections to the expression for $\ln R\left(t\right)$ but these are power-suppressed at small $t$:
\begin{equation}
\begin{aligned}
\tilde{L}\left(t\right)&=\ln\left(\frac{1}{t}\right)+\ln\left(1-\frac{t}{t_{max}}+t\right)
\\
&=L\left(t\right)+\left(t-\frac{t}{t_{max}}\right)-\frac{1}{2}\left(t-\frac{t}{t_{max}}\right)^2+\ldots,
\end{aligned}
\label{eqfoureighteen}
\end{equation}
and so $\tilde{L}\left(t\right)\rightarrow L\left(t\right)$ in the important limit $t\rightarrow0$.

\section{Results of NNLO+NLL matching}
\label{sec:results}
To perform the matching, the \textit{integrated} perturbation series
coefficients are required as in Eq.~(\ref{eqfoureleven}). For $R_1\left(t\right)$, the
analytic result is
\begin{equation}
\begin{aligned}
R_1\left(t\right)=&-\frac{8}{3}\ln^2\left(\frac{t}{1-t}\right)-4\left(1-2t\right)\ln\left(\frac{t}{1-2t}\right)+\frac{4\pi^2}{9}\\&-\frac{10}{3}+8t+6t^2-\frac{16}{3}\text{Li}_2\left(\frac{t}{1-t}\right),
\end{aligned}
\label{eqfivetwo}
\end{equation}
where
\begin{equation}
\text{Li}_2\left(z\right)\equiv\int^0_zdx\frac{\ln\left(1-x\right)}{x}
\label{eqfivethree}
\end{equation}
is the dilogarithm function. $R_2\left(t\right)$ and $R_3\left(t\right)$ were obtained by interpolating the differential results from \texttt{EERAD3} and then numerically integrating them. For $R_3\left(t\right)$, the \texttt{EERAD3} results were first smoothed by taking
\begin{equation}
\frac{dR_3}{dt}\left(t_i\right)\rightarrow\frac{1}{3}\left[\frac{dR_3}{dt}\left(t_{i+1}\right)+\frac{dR_3}{dt}\left(t_{i}\right)+\frac{dR_3}{dt}\left(t_{i-1}\right)\right],
\label{eqfivefour}
\end{equation}
repeatedly until a smooth curve was obtained. The peak near $t=0$ had to be reintroduced by hand, as this smoothing technique always results in the peak value being reduced.

$R\left(t\right)$ was computed to NNLO+NLL precision using
Eqs.~(\ref{eqfoureleven}) and (\ref{eqfoureighteen})
with $t_{max}=0.42$ in $\tilde{L}$, as this is the maximum value of $t$
kinematically allowed in the five parton limit. The differential cross
section was then obtained by
numerically differentiating $R(t)$. The results at a range of
energies are shown by the red/darker curves in
Figs~\ref{fig:unshift1}-\ref{fig:unshift3}.
The values of $\alpha_s\left(Q\right)$
were calculated as described earlier for the unresummed NNLO (green/lighter)
curves. The shaded area around each line shows the
renormalisation scale uncertainty found by taking
$\mu_R^2\in\left[Q^2/2,2Q^2\right]$.
\subsection{Comparison with experimental data}  
\label{sec:comp}
The matched, resummed differential thrust distribution was
compared with data from a wide range of experiments, as listed in
Table~\ref{tab:data}. The points in
Figs.~\ref{fig:unshift1}-\ref{fig:unshift3}
show the data at an illustrative selection of energies.
The error bars represent the experimental
statistical and systematic errors, added in quadrature.

\begin{table}
\begin{center}
\begin{tabular}{|l|c|c|r|r|}
\hline
Experiment & $Q$/GeV & Ref. & No. Pts. & $\chi^2$ \\
\hline
TASSO & $14.0$ & \cite{tasso} & 4 & 8.2 \\
TASSO & $22.0$ & \cite{tasso} & 6 & 2.8 \\
TASSO    & $35.0$ & \cite{tasso} & 8 & 0.7 \\
JADE    & $35.0$ & \cite{jade} & 10 & 10.5 \\
L3      & $41.4$ & \cite{l3} & 8 & 3.4 \\
JADE & $44.0$ & \cite{jade} & 10 & 3.8 \\
TASSO & $44.0$ & \cite{tasso} & 8 & 6.8 \\
DELPHI & $45.0$ & \cite{delphi} & 11 & 11.6 \\
AMY & $54.5$ & \cite{amy} & 4 & 4.9 \\
L3 & $55.3$ & \cite{l3} & 8 & 3.2 \\
L3 & $65.4$ & \cite{l3} & 8 & 7.5 \\
DELPHI & $66.0$ & \cite{delphi} & 11 & 14.5 \\
L3 & $75.7$ & \cite{l3} & 8 & 1.9 \\
DELPHI & $76.0$ & \cite{delphi} & 11 & 10.3 \\
L3 & $82.3$ & \cite{l3} & 8 & 4.0 \\
L3 & $85.1$ & \cite{l3} & 8 & 3.6 \\
OPAL & $91.0$ & \cite{opal} & 5 & 11.9 \\
ALEPH & $91.2$ & \cite{aleph} & 27 & 16.1 \\
DELPHI & $91.2$ & \cite{delphi} & 11 & 18.8 \\
SLD & $91.2$ & \cite{sld} & 6 & 2.7 \\
L3 & $130.1$ & \cite{l3} & 10 & 14.6 \\
ALEPH & $133.0$ & \cite{aleph} & 6 & 7.2 \\
OPAL & $133.0$ & \cite{opal} & 5 & 6.5 \\
L3 & $136.1$ & \cite{l3} & 10 & 37.3 \\
ALEPH & $161.0$ & \cite{aleph} & 6 & 5.5 \\
L3 & $161.3$ & \cite{l3} & 10 & 4.0 \\
ALEPH & $172.0$ & \cite{aleph} & 6 & 14.0 \\
L3 & $172.3$ & \cite{l3} & 10 & 2.1 \\
OPAL & $177.0$ & \cite{opal} & 5 & 1.1 \\
L3 & $182.8$ & \cite{l3} & 10 & 2.7 \\
ALEPH & $183.0$ & \cite{aleph} & 6 & 4.0 \\
DELPHI & $183.0$ & \cite{delphi} & 13 & 33.1 \\
L3 & $188.6$ & \cite{l3} & 10 & 3.4 \\
ALEPH & $189.0$ & \cite{aleph} & 6 & 6.7 \\
DELPHI & $189.0$ & \cite{delphi} & 13 & 22.7 \\
DELPHI & $192.0$ & \cite{delphi} & 13 & 12.1 \\
L3 & $194.4$ & \cite{l3} & 10 & 1.2 \\
DELPHI & $196.0$ & \cite{delphi} & 13 & 39.7 \\
OPAL & $197.0$ & \cite{opal} & 5 & 10.0 \\
ALEPH & $200.0$ & \cite{aleph} & 6 & 21.0 \\
DELPHI & $200.0$ & \cite{delphi} & 13 & 7.1 \\
L3 & $200.0$ & \cite{l3} & 9 & 6.5 \\
DELPHI & $202.0$ & \cite{delphi} & 13 & 14.9 \\
DELPHI & $205.0$ & \cite{delphi} & 13 & 12.6 \\
ALEPH & $206.0$ & \cite{aleph} & 6 & 7.0 \\
L3 & $206.2$ & \cite{l3} & 10 & 10.0 \\
DELPHI & $207.0$ & \cite{delphi} & 13 & 11.7 \\

\hline
Total & \text{} & \text{} & 430 & 466.0 \\
\hline
\end{tabular}
\caption{Data sets used and best-fit $\chi^2$ contributions.\label{tab:data}}
\end{center}
\end{table}

There are a few features common to the graphs at all
energies. Firstly, the resummed distribution and the NNLO distribution
are almost identical away from the two-jet region. However, in this
low-$t$ limit the resummed distribution peaks, in line with the
experimental data, whereas the NNLO distribution carries on
increasing. Thus resummation has significantly improved the
theoretical prediction in the two-jet limit, as we had expected. 

It should be noted that the kink around $t=0.33$ in all of the curves
is due to the LO term vanishing here for kinematic reasons. One would
expect that with many higher-order perturbation theory terms taken
into account (i.e. more partons present in the final state), this
would gradually smoothen, in line with the experimental data.

At all energies, the overall shape of the theoretical distribution is
similar to that of the data, but is shifted to a lower value of
$t$. This apparent shift $\delta t$ has a clear energy dependence -- at
the upper end of the energy range considered here, the NNLO+NLL and
experimental distributions are fairly close and the shift $\delta t$
is a very small correction. On decreasing the energy, the shift
becomes more pronounced and at low energies the theoretical
distribution is clearly not consistent with the data. There is no
obvious way that this could be remedied by the inclusion of
sub-leading logarithms or higher fixed-order terms, and so we now turn
to considering \textit{non-perturbative} effects for an
explanation. The increasing discrepancy at low energies is also
consistent with this interpretation, as we expect such effects to have a
$1/Q$ dependence, as mentioned in Sect.~\ref{sec:intro}. To verify
that these discrepancies are due to non-perturbative effects, the
exact form of their energy dependence was investigated.

\subsection{Power dependence of discrepancies}
\label{sec:power}
As both the experimental data and \texttt{EERAD3} results are given as
histograms, and not in terms of individual values of $t$, the
integrated thrust distribution $R\left(t\right)$ should be
slightly more accurate than $d\sigma/dt$ as it does not involve the
assumption of a uniform distribution over the width of each histogram
bin $\Delta t$.

\begin{figure}
\begin{center}
\includegraphics[scale=0.51]{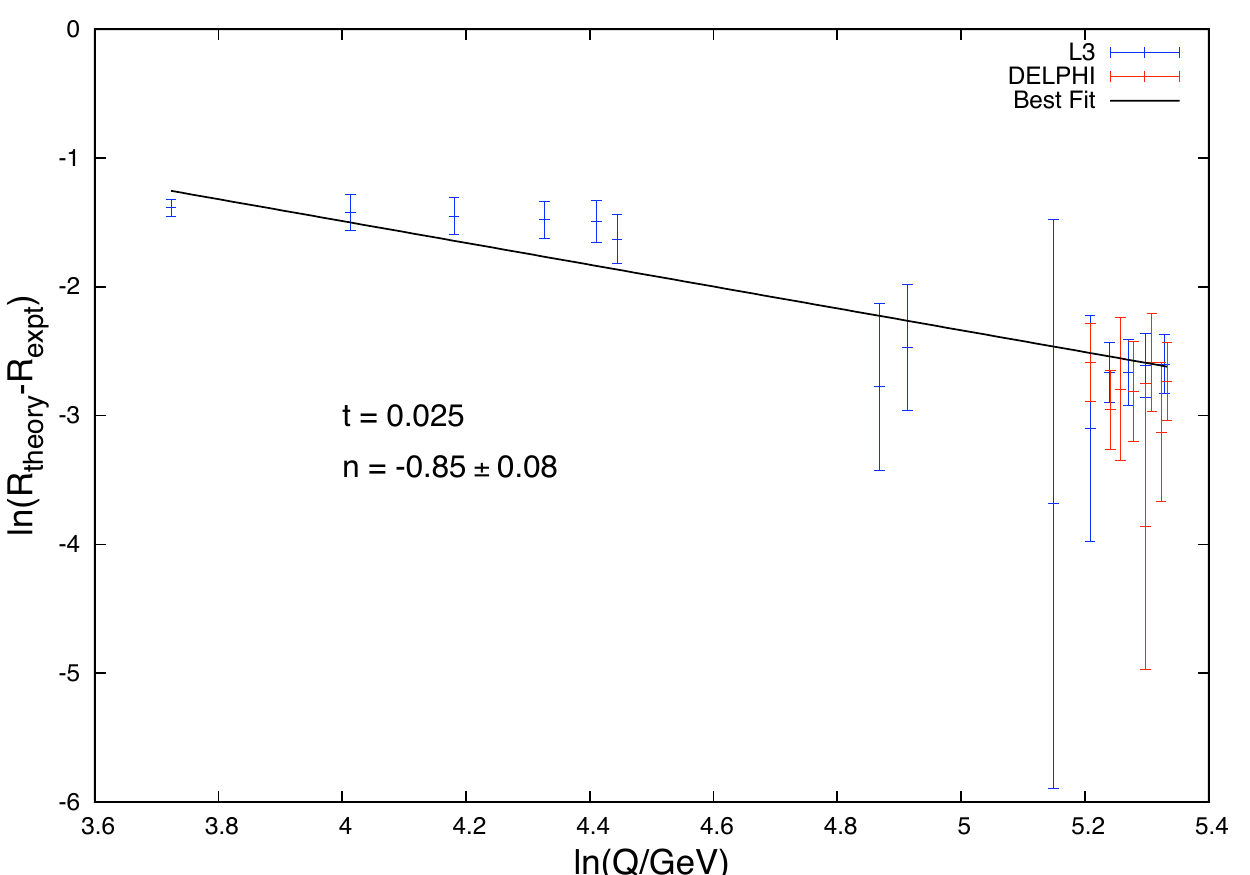}
\includegraphics[scale=0.51]{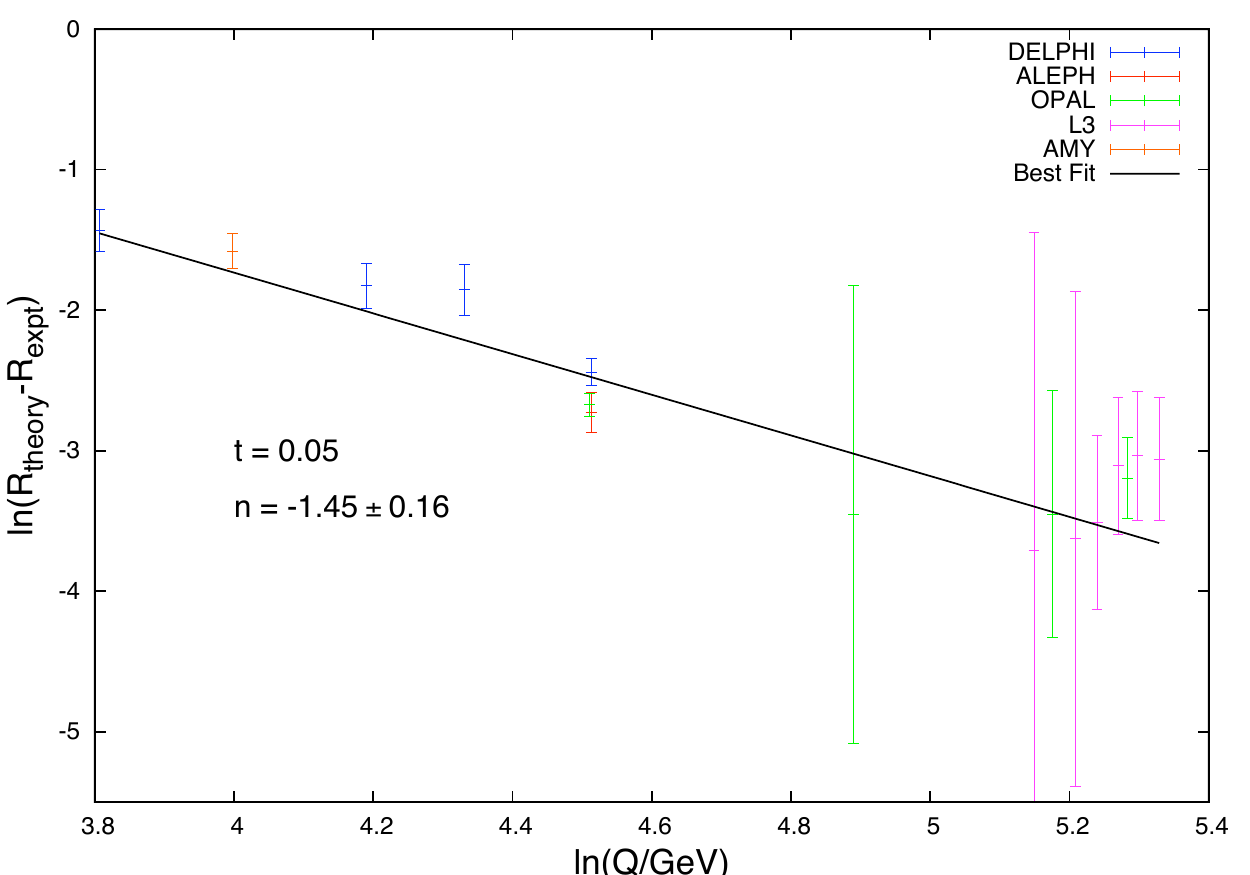}
\includegraphics[scale=0.51]{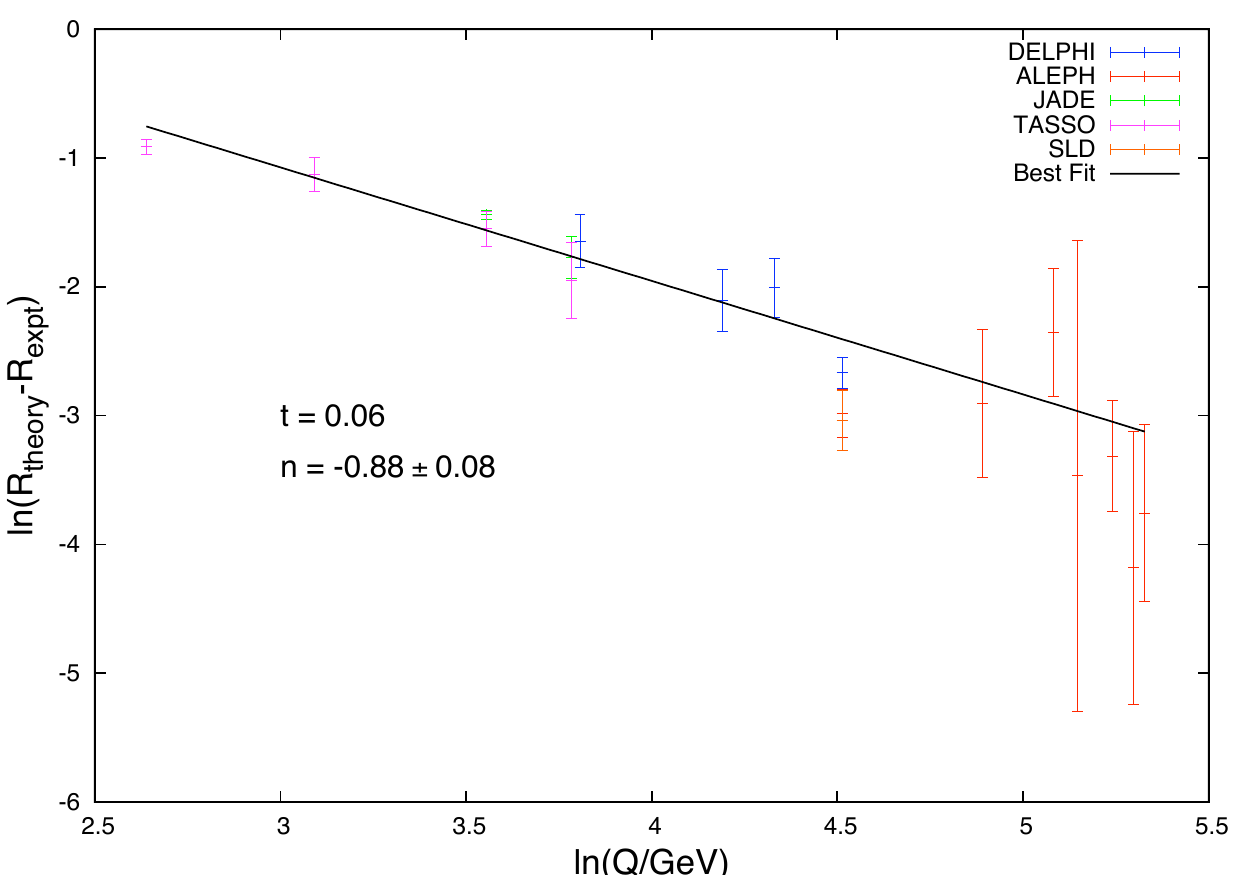}
\includegraphics[scale=0.51]{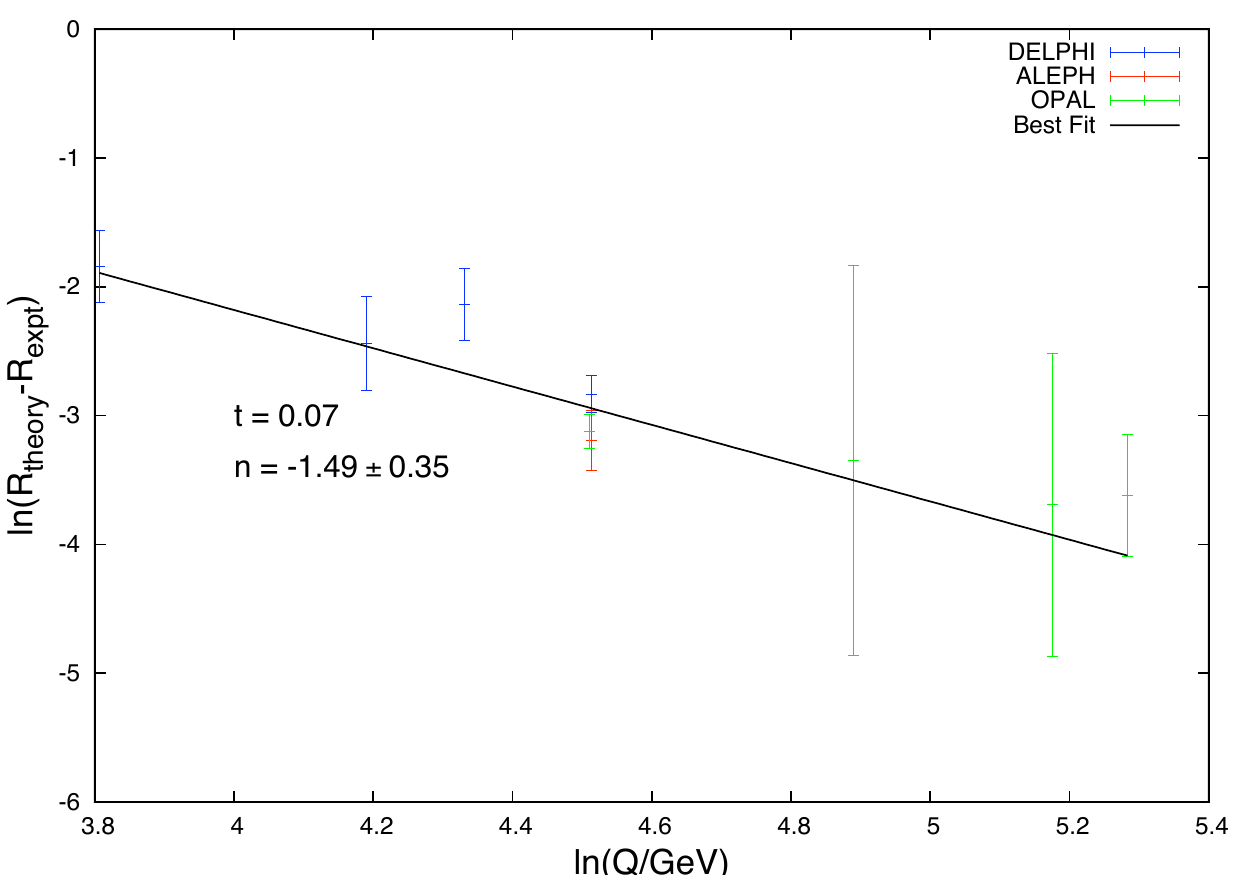}
\includegraphics[scale=0.51]{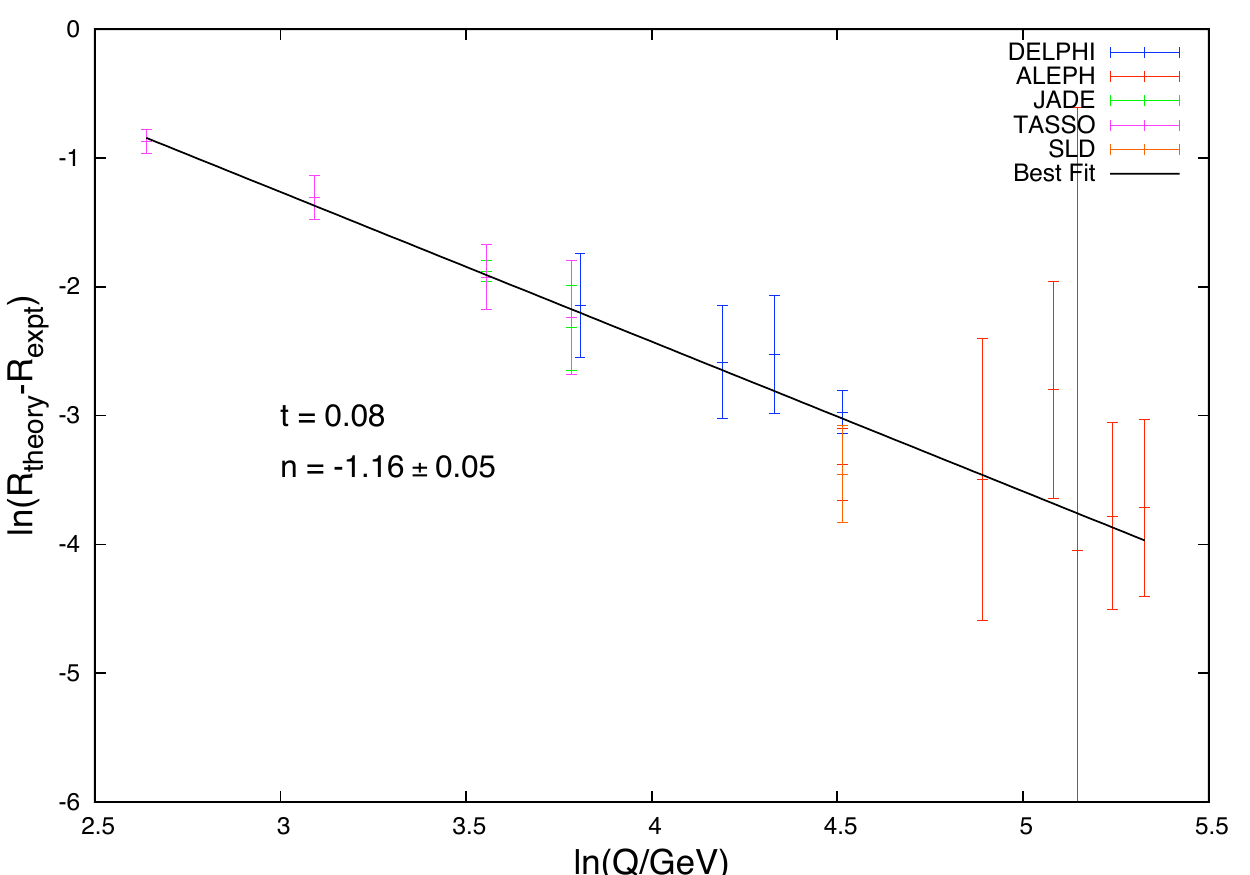}
\caption{Power dependence of corrections required to resolve
  theory/data discrepancy: $t=0.025-0.08$.\label{fig:log1}}
\end{center}
\end{figure}

\begin{figure}
\begin{center}
\includegraphics[scale=0.51]{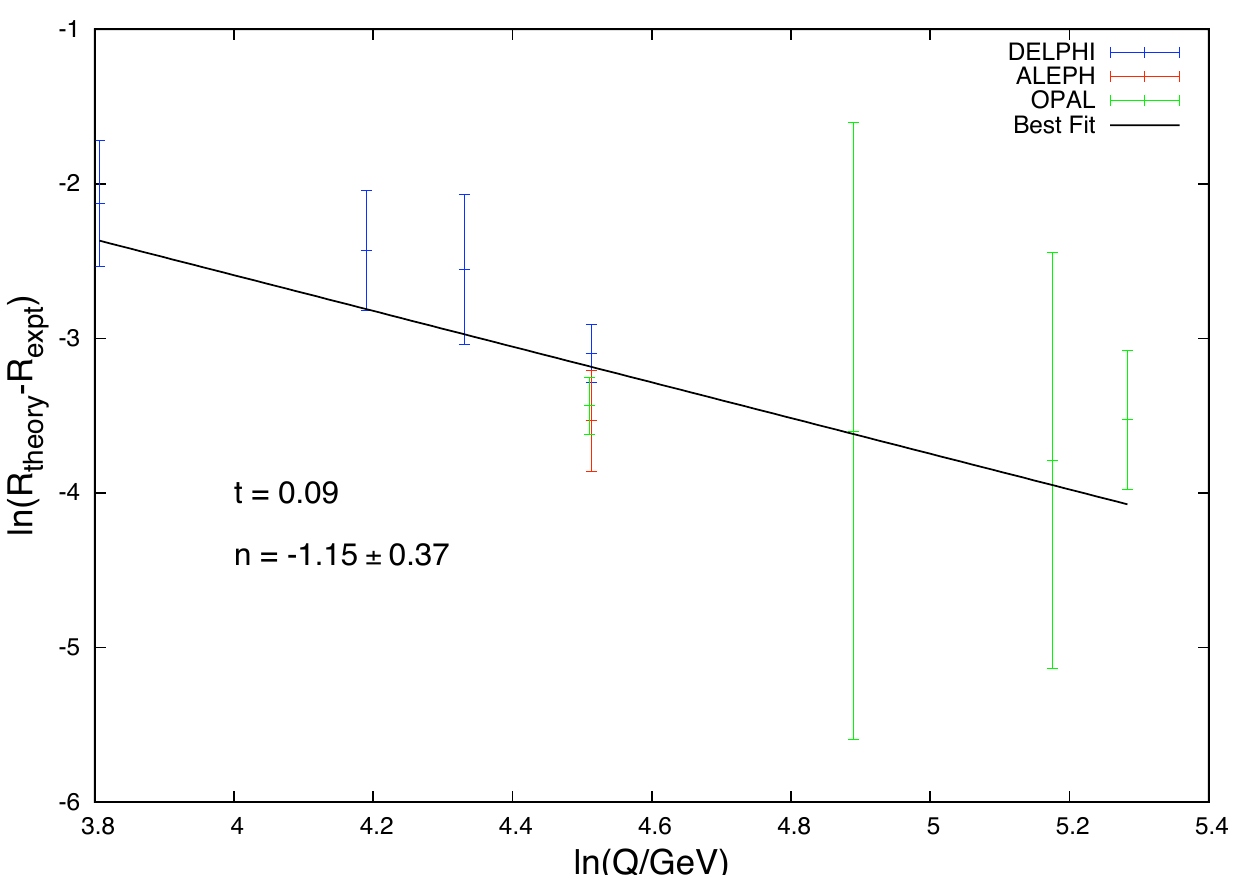}
\includegraphics[scale=0.51]{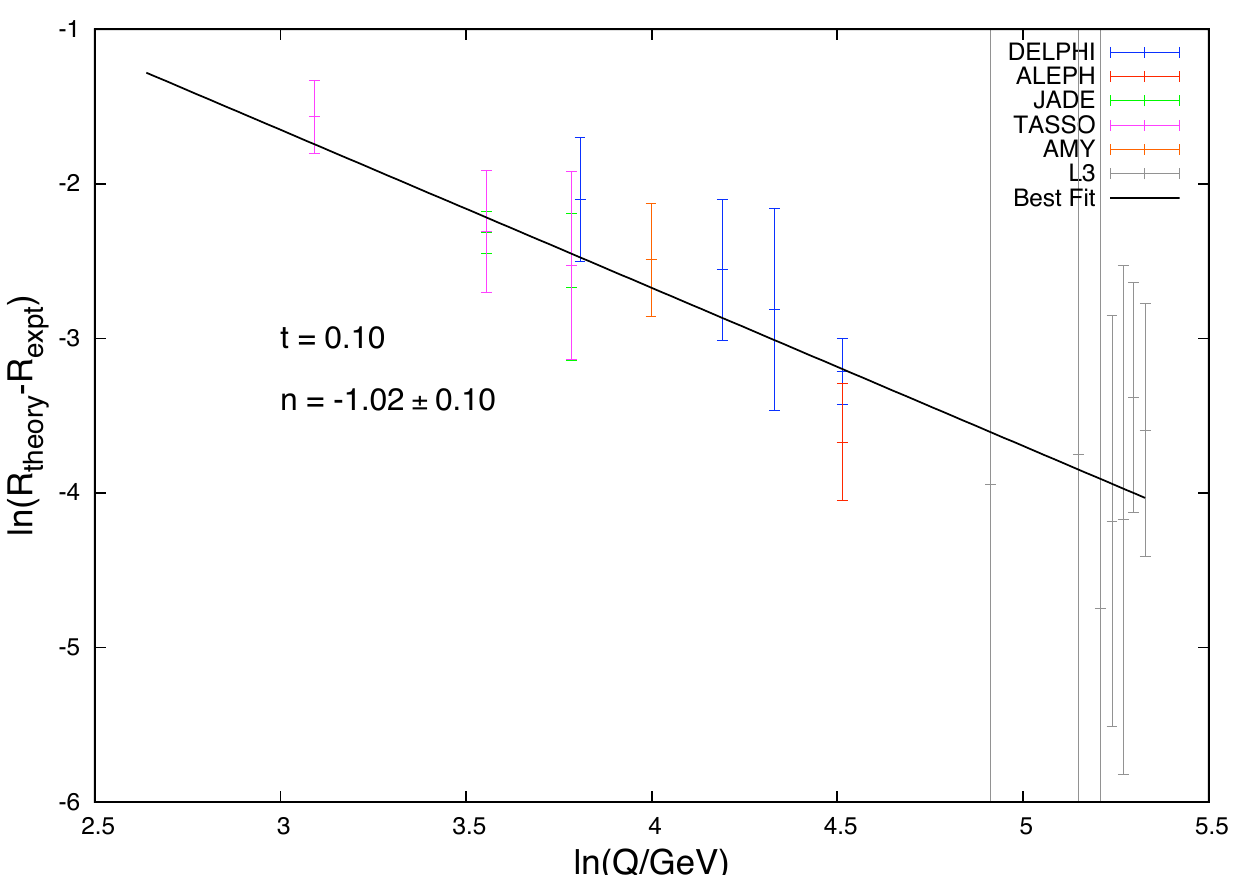}
\includegraphics[scale=0.51]{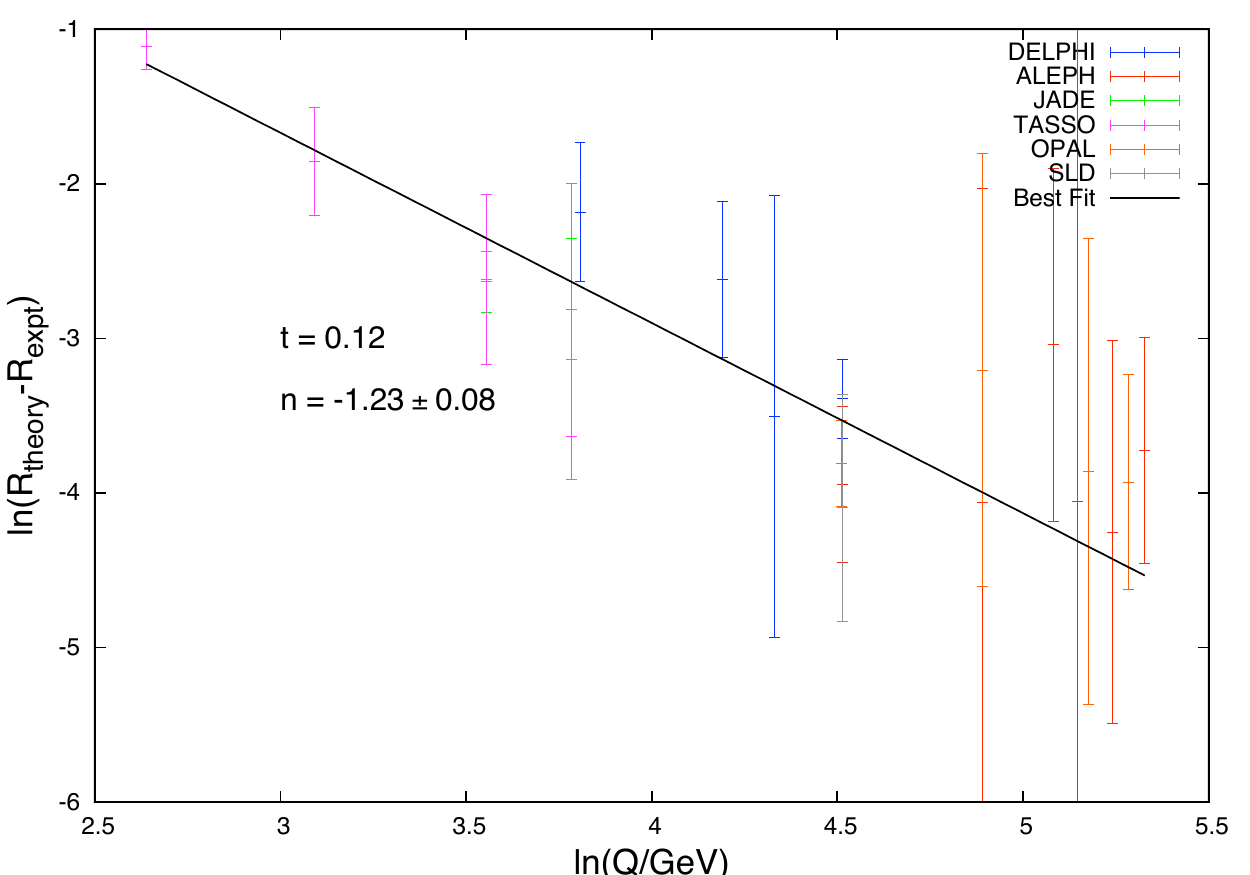}
\includegraphics[scale=0.51]{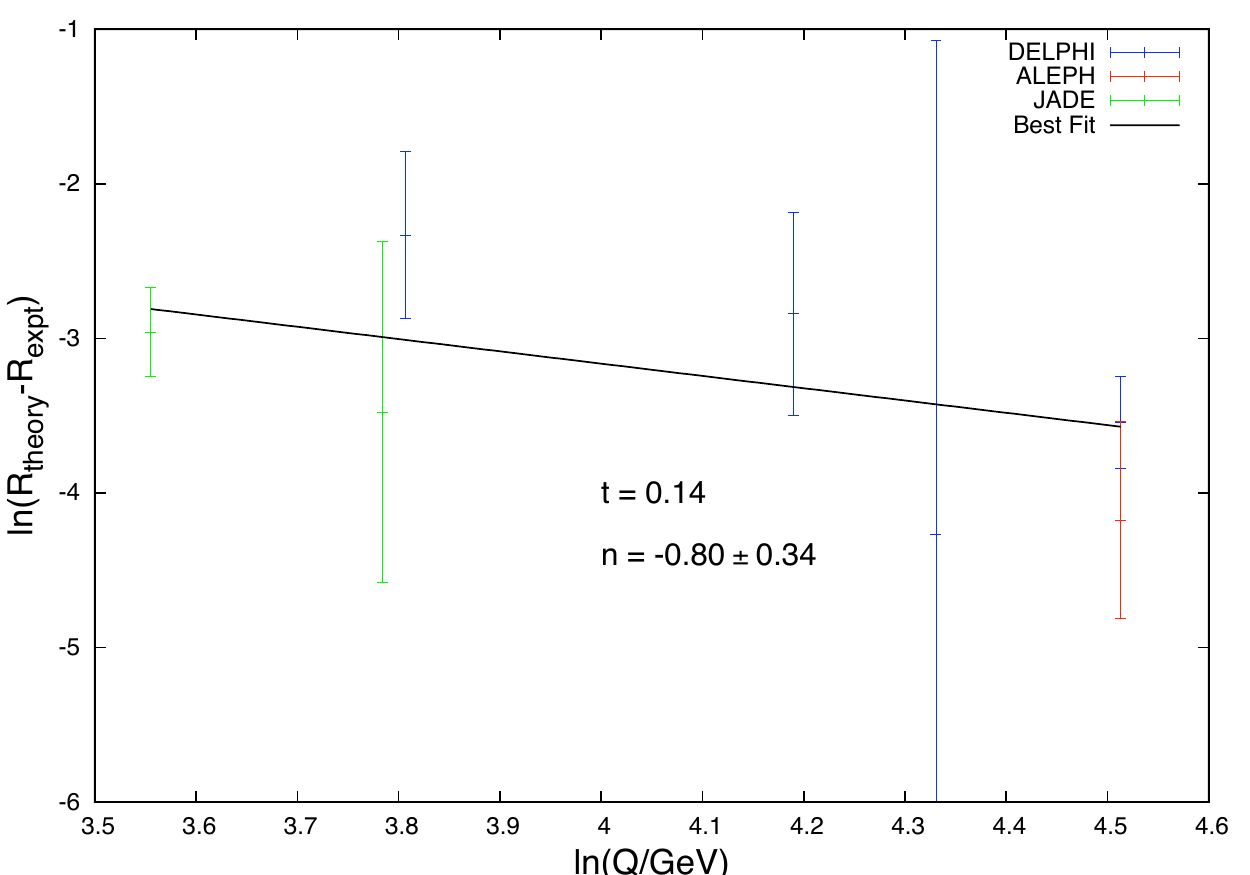}
\includegraphics[scale=0.51]{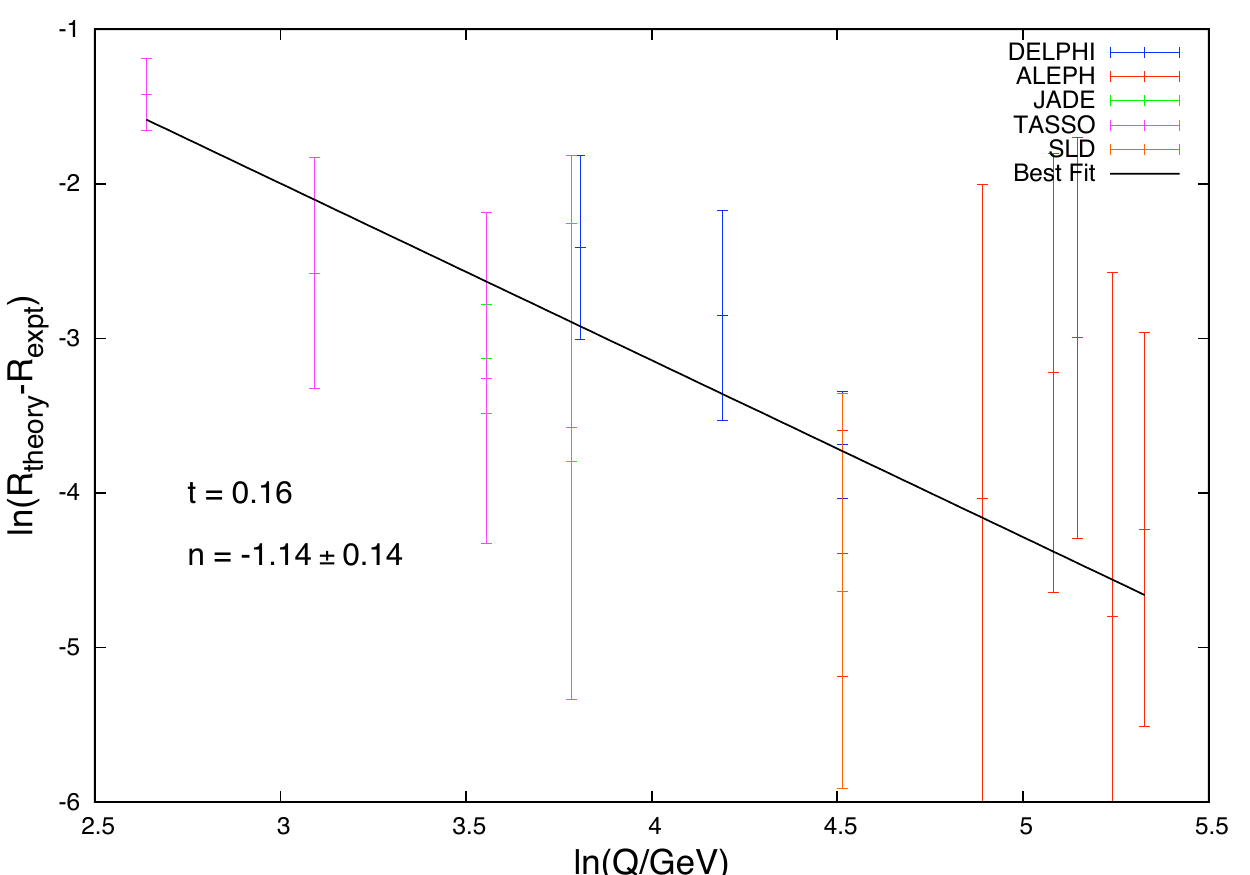}
\caption{Power dependence of corrections required to resolve
  theory/data discrepancy: $t=0.09-0.16$.\label{fig:log2}}
\end{center}
\end{figure}

\begin{figure}
\begin{center}
\includegraphics[scale=0.51]{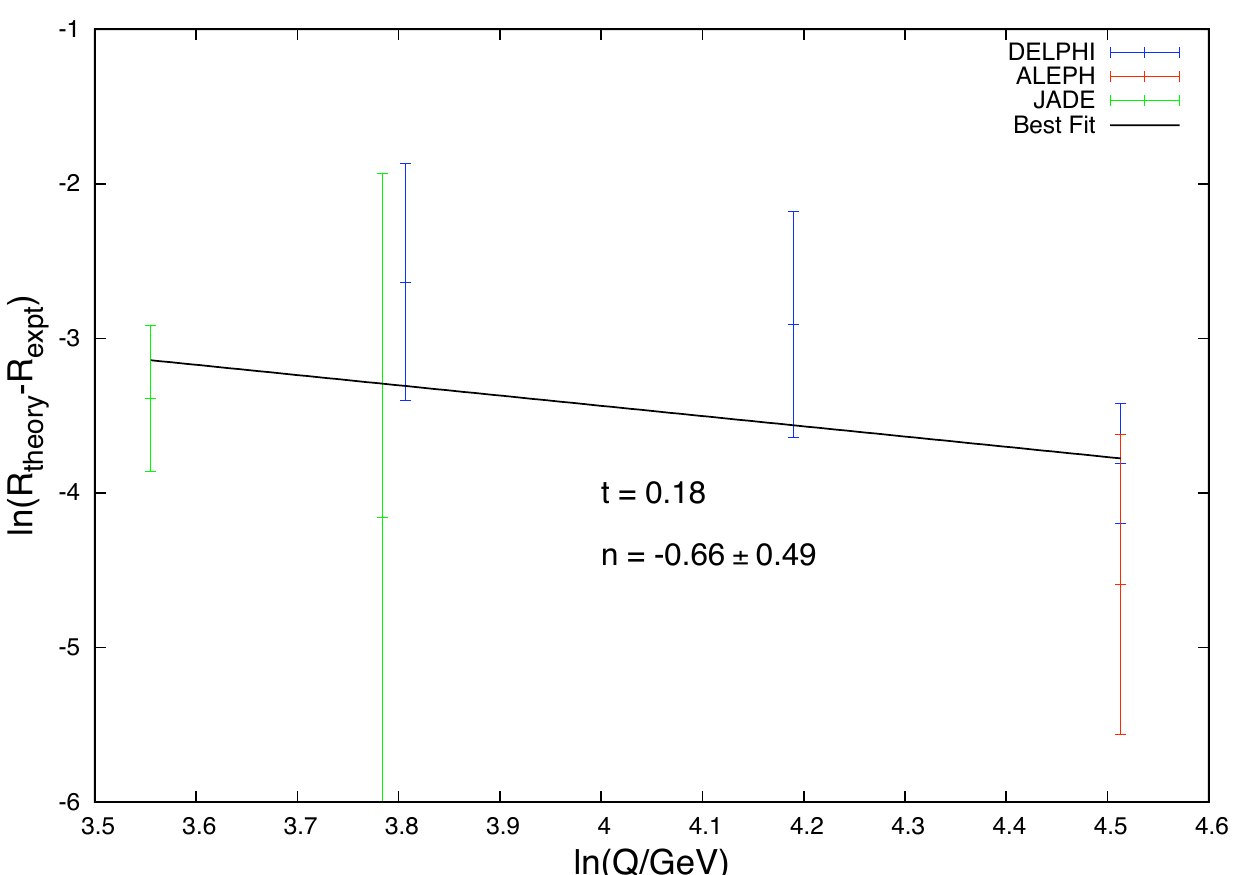}
\includegraphics[scale=0.51]{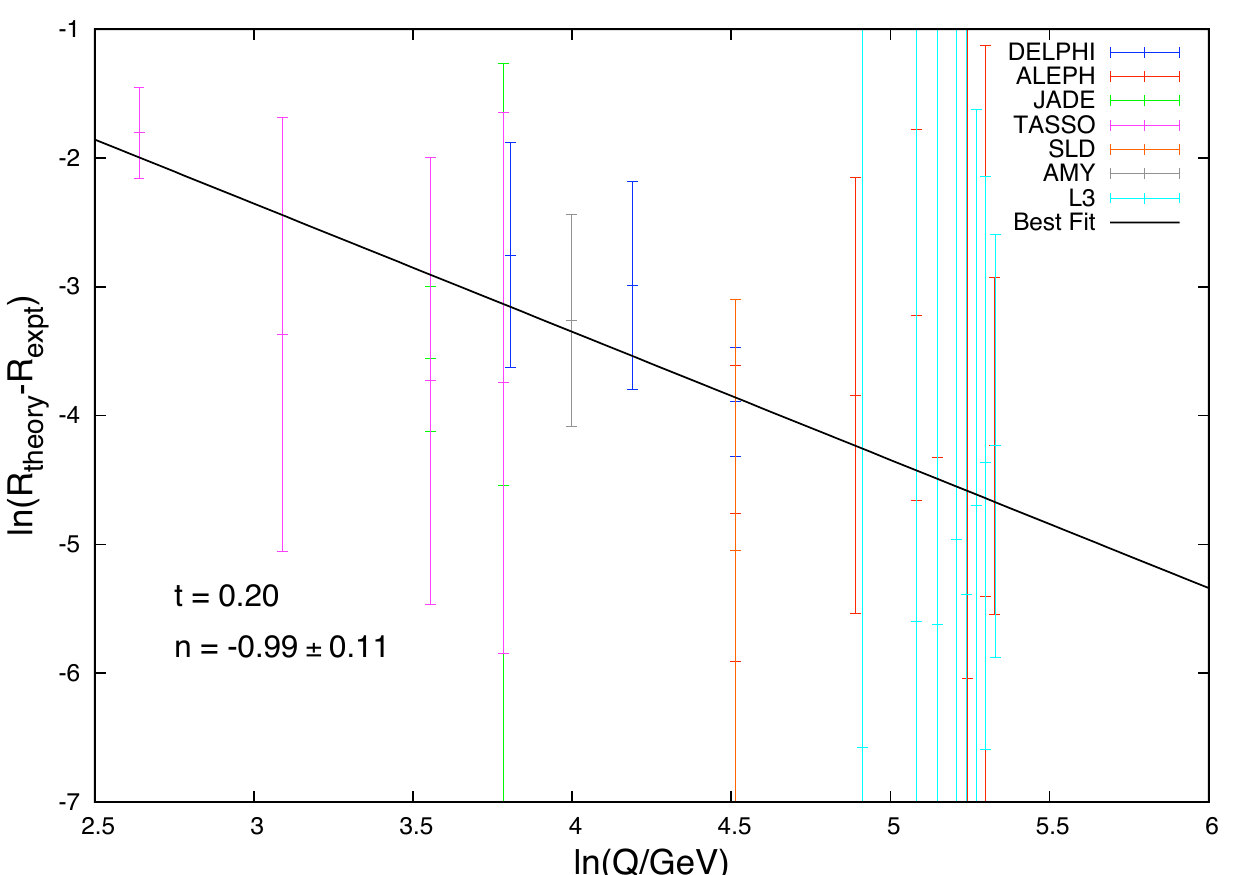}
\includegraphics[scale=0.51]{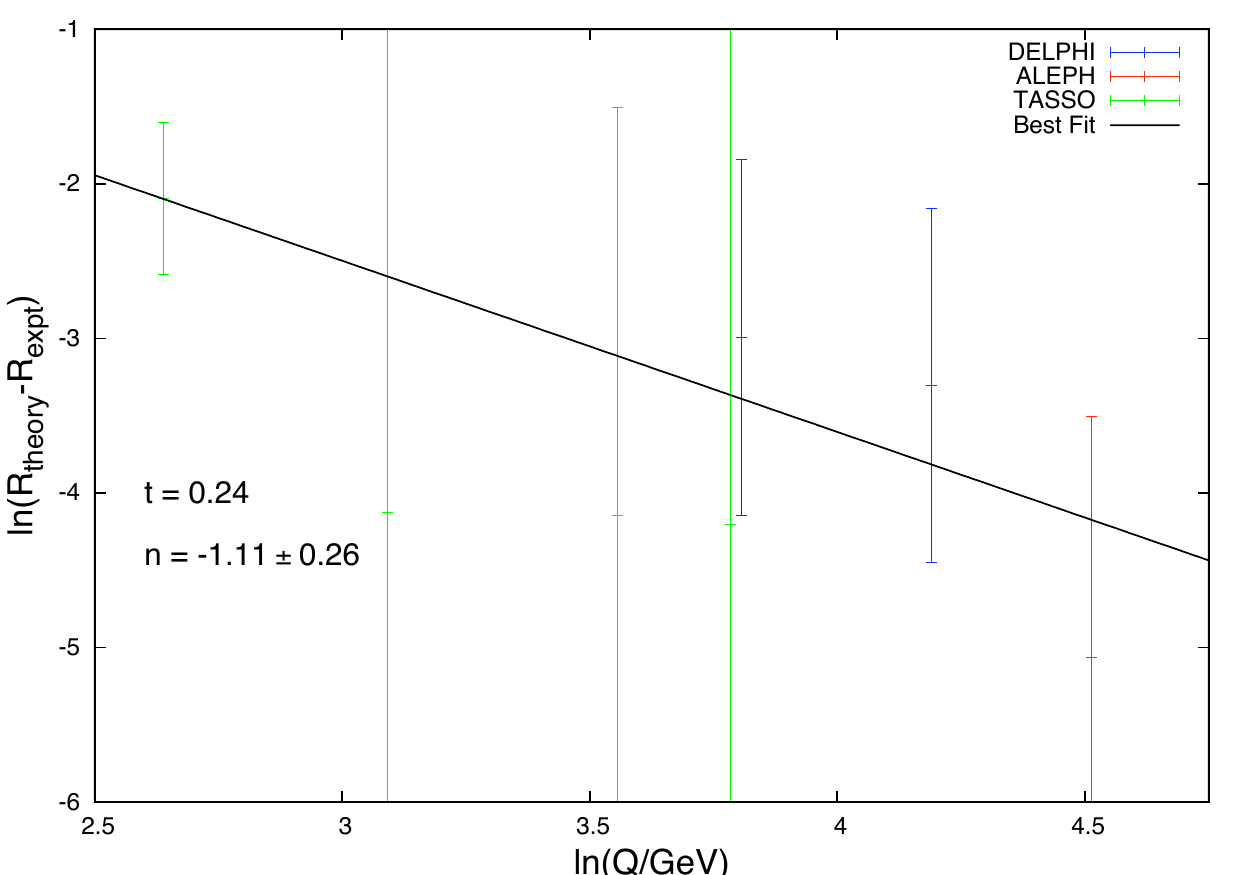}
\caption{Power dependence of corrections required to resolve
  theory/data discrepancy: $t=0.18-0.24$.\label{fig:log3}}
\end{center}
\end{figure}

Graphs of $\ln\left(R_{\text{theory}}-R_{\text{expt}}\right)$ against
$\ln\left(Q/\text{GeV}\right)$ were plotted for $0.025\leq t\leq0.24$
and, anticipating corrections proportional to an inverse power of $Q$,
a linear fit was made to each plot such that the gradient $n$ of the
straight line gives the power dependence of the required correction
($\propto Q^{n}$). The results are shown in
Figs.~\ref{fig:log1}-\ref{fig:log3}. This $t$ range was chosen since
at lower values of $t$ there is no obvious straight line (due to the
distributions peaking), and at higher values of $t$ the percentage
errors on the gradient become large due to
$R_{\text{theory}}-R_{\text{expt}}$ quickly decreasing to zero (as
both normalised cross-sections converge to
1).

Although not totally conclusive, these results are consistent with
power corrections of the form $1/Q$, and we turn now to considering
the quantitative form of these non-perturbative corrections to the
thrust distribution.

\section{Non-perturbative corrections}
\label{sec:NP}
\subsection{The low-scale effective coupling}
\label{sec:effective}
Although there are various ways to phenomenologically treat
non-perturbative effects in QCD, one of the most intuitive is by means
of a \textit{low-scale effective coupling} \cite{one}. In this
approach, the running coupling (\ref{eqtwosix}) is replaced by an
effective coupling $\alpha_{\text{eff}}\left(\mu\right)$, which
differs from the standard perturbative $\alpha_s(\mu)$ in the
infra-red region where the latter diverges. Using this finite
effective coupling allows us to use the formalism of perturbation
theory to describe non-perturbative effects which cannot be probed
using standard perturbative QCD.

Various forms for $\alpha_{\text{eff}}(\mu)$ have been
proposed~\cite{Solovtsov:1997at,nine} that have high-energy behaviour
consistent with $\alpha_s$, but we will not be concerned with their
details here. The only parameter we require is the `average' value of
the effective coupling below the infra-red matching scale $\mu_I$
where $\alpha_s$ and $\alpha_{\text{{eff}}}$ begin to differ:
\begin{equation}
\alpha_0\left(\mu_I\right)\equiv\frac{1}{\mu_I}\int_0^{\mu_I}d\mu\text{ }\alpha_{\text{eff}}\left(\mu\right).
\label{eqsixone}
\end{equation}
We make the additional assumption that $\alpha_{\text{eff}}$ is small
enough in the infra-red region that we can neglect terms of order
$\alpha_{\text{eff}}^2$ and higher.

\subsection{Non-perturbative shift in thrust distribution}
\label{sec:NPthrust}
In deriving the form of the NNLO+NLL prediction used earlier, the low
$\mu$ region in Eq.~(\ref{eqfourfive}) was neglected as it produced a
subleading contribution. We now include this region by subtracting the
fixed-order NNLO contribution from $\mu\le\mu_I$ and replacing it
with a contribution due to the effective coupling. We are thus
removing the renormalon contributions to the perturbation series (up
to NNLO) and incorporating all $1/Q$-dependent behaviour into
$\alpha_{\text{eff}}$. 

Firstly, we note that the order of integration in (\ref{eqfourfive})
can be changed, to give
\begin{eqnarray}
\ln\tilde{J}_{\nu}^{\mu}\left(Q^2\right)&=&\frac{2C_F}{\pi}\int_0^{Q}\frac{d\mu}{\mu}\alpha_s\left(\mu\right)\left(1-K\frac{\alpha_s\left(\mu\right)}{2\pi}\right)^{-1}\nonumber\\
&&\int_{\frac{\mu^2}{Q^2}}^{\frac{\mu}{Q}}\frac{du}{u}\left(e^{-u\nu Q^2}-1\right)\;.
\label{eqsixfour}
\end{eqnarray}
Inserting the NNLO perturbative running coupling
\begin{eqnarray}
\alpha_s\left(\mu\right)&=&\alpha_s\left(\mu_{R}\right)+\alpha_s^2\left(\mu_{R}\right)\frac{\beta_0}{\pi}\ln\frac{\mu_{R}}{\mu}\nonumber\\
&+&\alpha_s^3\left(\mu_{R}\right)\left[\left(\frac{\beta_0}{\pi}\right)^2\ln^2\frac{\mu_{R}}{\mu}+\frac{\beta_1}{2\pi^2}\ln\frac{\mu_{R}}{\mu}\right],
\label{eqsixfive}
\end{eqnarray}
expanding the exponential to first order\footnote{Higher-order terms
  in the expansion would give corrections of order $1/Q^2$, which we
  neglect.}
 and integrating over the range $0\leq\mu\leq\mu_I$ gives an NNLO contribution of 
\begin{equation}
\begin{aligned}
&-\frac{2C_F}{\pi}\frac{\mu_I}{Q}\Biggl\{\alpha_s\left(\mu_R\right)+\alpha_s^2\left(\mu_R\right)\frac{\beta_0}{\pi}\left(\ln\frac{\mu_R}{\mu_I}+\frac{K}{2\beta_0}+1\right)\\
&+\alpha_s^3\left(\mu_R\right)\left(\frac{\beta_0}{\pi}\right)^2\biggl[\ln^2\frac{\mu_R}{\mu_I}+\left(\ln\frac{\mu_R}{\mu_I}+1\right)\\
&\left(2+\frac{\beta_1}{2\beta_0^2}+\frac{K}{\beta_0}\right)
+\frac{K^2}{4\beta_0^2}\biggr]\Biggr\}\nu Q^2.
\end{aligned}
\label{eqsixsix}
\end{equation}

It should be noted that $t$ is the conjugate variable to $\nu Q^2$ in
the Laplace transform (\ref{eqfourfour}) and thus the first-order
expansion of the exponential will only be a valid approximation in the
limit $t\gg\mu_I/Q$. Below this, we would need to retain higher order
terms in the expansion, which would require us to have a specific form
for $\alpha_{\text{eff}}\left(\mu\right)$.

Following a similar procedure with
$\alpha_{\text{eff}}\left(\mu\right)$ in the place of
$\alpha_s\left(\mu\right)$ gives a non-perturbative contribution of
\begin{equation}
-\frac{2C_F}{\pi}\int_0^{\mu_I}d\mu\text{ }\alpha_{\text{eff}}\left(\mu\right)\nu Q\equiv-\frac{2C_F}{\pi}\frac{\mu_I}{Q}\alpha_0\left(\mu_I\right)\nu Q^2,
\label{eqsixseven}
\end{equation}
where we have neglected terms of order $\alpha^2_{\text{eff}}$ as
previously noted.

By adding this, after subtracting the perturbative contribution
(\ref{eqsixsix}), we obtain the change in the quark jet
mass distribution caused by changing from a perturbative to an
effective coupling in the low-scale region below $\mu_I$.

Substituting the result into Eq.~(\ref{eqfourfour}), we see that the
effect of this non-perturbative contribution is to shift the thrust
distribution by an amount $\delta t$, such that
\begin{equation}
\left.\frac{1}{\sigma}\frac{d\sigma}{dt}\right|_t=\left.\left(\frac{1}{\sigma}\frac{d\sigma}{dt}\right)^{\text{pert.}}\right|_{t+\delta t},
\label{eqsixnine}
\end{equation}
where
\begin{equation}
\begin{aligned}
\delta t=&-\frac{4C_F}{\pi}\frac{\mu_I}{Q}\Biggl\{\alpha_0\left(\mu_I\right)-\alpha_s\left(\mu_R\right)\\
&-\alpha_s^2\left(\mu_R\right)\frac{\beta_0}{\pi}\left(\ln\frac{\mu_R}{\mu_I}+\frac{K}{2\beta_0}+1\right)\\
&-\alpha_s^3\left(\mu_R\right)\left(\frac{\beta_0}{\pi}\right)^2
\biggl[\ln^2\frac{\mu_R}{\mu_I}+\left(\ln\frac{\mu_R}{\mu_I}+1\right)\\
&\left(2+\frac{\beta_1}{2\beta_0^2}+\frac{K}{\beta_0}\right)+\frac{K^2}{4\beta_0^2}\biggr]\Biggr\},
\end{aligned}
\label{eqsixten}
\end{equation}
to NNLO. This $1/Q$-dependent shift is precisely what is required to
account for the differences between the perturbative and experimental
distributions seen in Sect.~\ref{sec:results}.
 
\subsection{\boldmath Determination of $\alpha_s$ and $\alpha_0$}
\label{sec:determ}
By applying the shift (\ref{eqsixten}) to the perturbative results, we
expect to reduce significantly the differences between the theoretical
and experimental distributions. Comparison of these differences to the
predicted form of $\delta t$ allows us to estimate the values of
$\alpha_0$ and $\alpha_s$.

Maximum accuracy was obtained by comparing the experimental
distribution with a discretely-defined theoretical distribution
\begin{equation}
\frac{R\left(t+\Delta t\right)-R\left(t\right)}{\Delta t},
\label{eqsixeleven}
\end{equation}
where $\Delta t$ is the bin width of the experimental data.

The NNLO+NLL+shift distribution was calculated as a function of $\alpha_0$ and
$\Lambda_{\overline{MS}}^{\left(5\right)}$.
This calculation was performed for $0.05\leq t\leq 0.33$, at the centre-of-mass
energies listed previously in Table~\ref{tab:data} (i.e.\ in the range
$14$ GeV$\leq Q\leq 207$ GeV). $\chi^2$ was calculated for each pair
of input parameters, with its minimum corresponding to the best-fit values.   

The upper limit for the fits was chosen as $t=0.33$ since the
difference between the theoretical and experimental distributions
above this value is largely
due to the small number of final state partons in the theoretical
calculation, as previously explained, rather than to any
non-perturbative effects. We noted previously that the
non-perturbative results are strictly valid only in the range $t\gg\mu_I/Q$;
in fact we found satisfactory fits using an energy-dependent lower cut-off
$t\geq\max\{\mu_I/Q,0.05\}$.

For infra-red matching scale $\mu_I=2$ GeV, best fit values of 
\begin{equation}
\begin{aligned}
\alpha_0\left(2\text{ GeV}\right)&=0.59\pm 0.03\;,\\
\Lambda_{\overline{MS}}^{\left(5\right)}&=0.190^{+0.025}_{-0.022}\text{ GeV}
\end{aligned}
\label{eqsixtwelve}
\end{equation}
were obtained, with $\chi^2/\text{d.o.f.}=466.0/428\approx 1.09$.
The quoted errors correspond to one standard deviation, computed as
recommended by the Particle Data Group~\cite{Amsler:2008zz}: the value of
$\chi^2$ corresponding to the $1\sigma$ (68.3\% C.L.) contour
was rescaled by the value of $\chi^2/\text{d.o.f.}$, giving
$\chi^2=480.6$, i.e.\ $\Delta\chi^2=14.6$.

The contribution to $\chi^2$ from each data set is shown in
Table~\ref{tab:data}. It should be noted that the few data sets with
$\chi^2/\text{no. pts.}\gg1$ are not generally inconsistent with the
shifted distribution, but simply have a few outlying points giving a
large contribution.

\begin{figure}
\begin{center}
\includegraphics[scale=0.6]{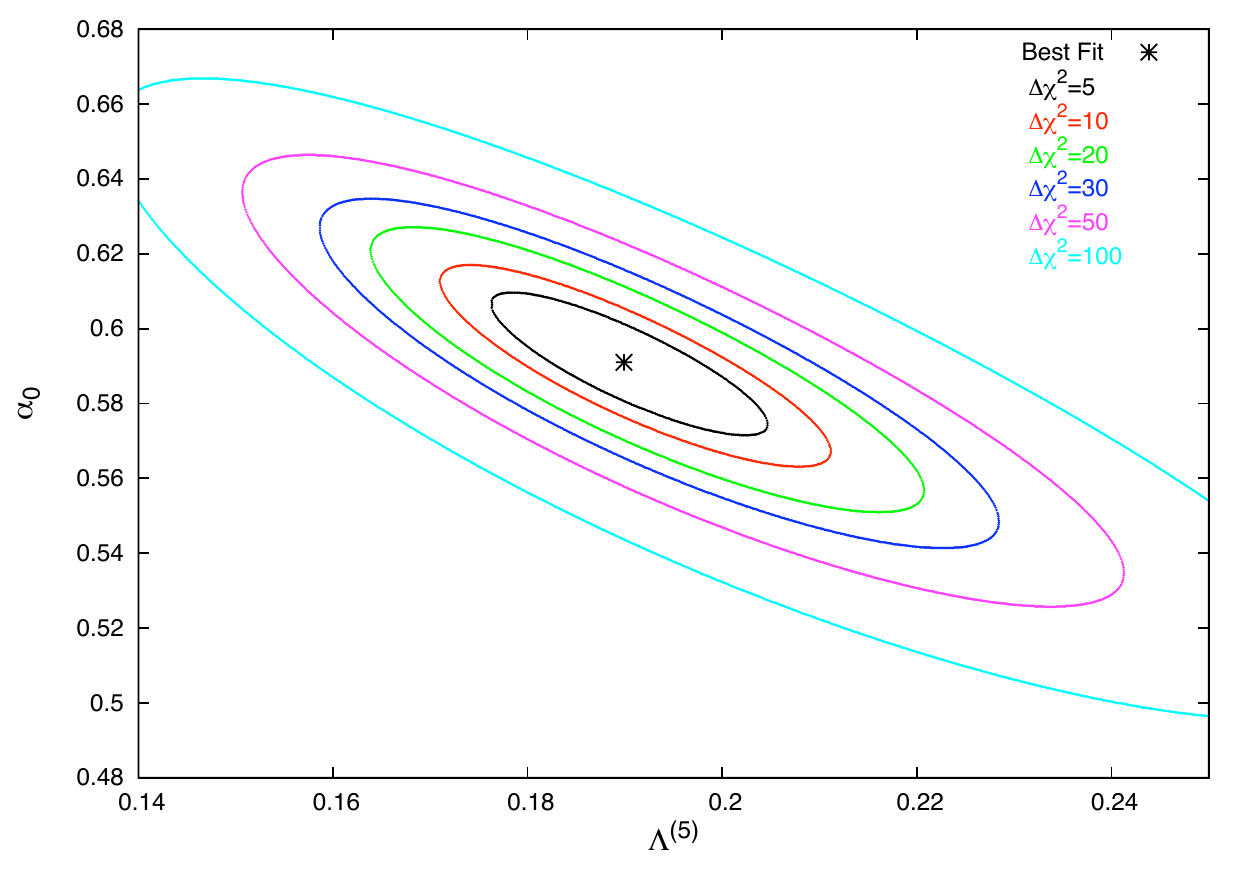}
\caption{$\chi^2$ contour plot in
$\left(\Lambda_{\overline{MS}}^{\left(5\right)}\text{ , }\alpha_0\right)$ space.\label{fig:chisq}}
\end{center}
\end{figure}

The contour plot in Fig.~\ref{fig:chisq} shows the ranges of
$\alpha_0$ and $\Lambda_{\overline{MS}}^{\left(5\right)}$ which give
fits within $\Delta\chi^2$ of the best-fit value of $\chi^2$, and
also demonstrates the correlation between these two parameters.

Varying the renormalisation scale $\mu_{R}^{2}\in\left[Q^2/2,2Q^2\right]$
gave best fit values in the range $\alpha_0\left(2\text{ GeV}\right)=0.585$,
$\Lambda_{\overline{MS}}^{\left(5\right)}=0.173$ GeV to
$\alpha_0\left(2\text{ GeV}\right)=0.598$,
$\Lambda_{\overline{MS}}^{\left(5\right)}=0.210$ GeV with no
significant change in the quality of fit. Thus we find
\begin{equation}
\Lambda_{\overline{MS}}^{\left(5\right)}=0.190^{+0.025+0.020}_{-0.022-0.017}\text{ GeV}
\end{equation}
where the first error is the combined experimental statistical and systematic
error and the second is due to the theoretical renormalisation scale
uncertainty. The corresponding strong coupling constant is
\begin{equation}
\alpha_s\left(91.2\text{ GeV}\right)=0.1164^{+0.0022+0.0017}_{-0.0021-0.0016}\;,
\label{eqsixthirteen}
\end{equation}
or, combining all the errors in quadrature,
\begin{equation}
\alpha_s\left(91.2\text{ GeV}\right)=0.1164^{+0.0028}_{-0.0026}\;,
\label{eqsixthirteen}
\end{equation}
in good agreement with the world average value of
0.1176~\cite{Amsler:2008zz}.

To assess the importance of the NNLO terms, the analysis was repeated with
all those terms omitted, i.e.\  combining NLO+NLL in perturbation
theory with Eq.~(\ref{eqsixten}) without the ${\cal O}(\alpha_s^3)$
contribution.  The resulting best fit values were
\begin{equation}
\begin{aligned}
\alpha_0\left(2\text{ GeV}\right)&=0.51\pm 0.04\;,\\
\Lambda_{\overline{MS}}^{\left(5\right)}&=0.214^{+0.032+0.034}_{-0.027-0.026}\text{ GeV}\;,\\
\alpha_s\left(91.2\text{ GeV}\right)&=0.1185^{+0.0025+0.0027}_{-0.0024-0.0023}
\end{aligned}
\label{eqsixtwelve}
\end{equation}
with $\chi^2/\text{d.o.f.}=515.1/428\approx 1.20$.  Thus the NLO and
NNLO results are consistent but the inclusion of NNLO terms
consistently in both the perturbative prediction and the power correction
improves the quality of the fit and reduces the errors.

The most complete previous NLO study along similar
lines~\cite{MovillaFernandez:2001ed}, combining NLO+NLL in
perturbation theory with the NLO equivalent of Eq.~(\ref{eqsixten})
and covering a variety of event shapes but a slightly narrower range
of energies than that used here, obtained the
overall best fit at 
\begin{equation}
\begin{aligned}
\alpha_s\left(91.2\text{ GeV}\right)&=0.1171^{+0.0032}_{-0.0020},\\
\alpha_0\left(2\text{ GeV}\right)&=0.513^{+0.066}_{-0.045}
\end{aligned}
\label{eqmovilla}
\end{equation}
in good agreement with our results.  Their fit to the thrust
distribution alone gave
\begin{equation}
\begin{aligned}
\alpha_s\left(91.2\text{ GeV}\right)&=0.1173^{+0.0063}_{-0.0051},\\
\alpha_0\left(2\text{ GeV}\right)&=0.492^{+0.084}_{-0.070}
\end{aligned}
\label{eqmovilla}
\end{equation}
also in good agreement.

In the recent NNLO analysis \cite{eleven}, a range of event shapes
at energies at and above 91.2 GeV were fitted without resummation;
non-perturbative effects were estimated using Monte Carlo event generators. The
value obtained for the strong coupling was
$\alpha_s\left(91.2\text{ GeV}\right)=0.1240\pm 0.0033$.

To estimate the dependence of our results upon the
infra-red matching scale, a fit with $\mu_I=3$ GeV was made, yielding
$\alpha_0\left(3\text{ GeV}\right)=0.458\pm 0.025$ and
$\Lambda_{\overline{MS}}^{\left(5\right)}=0.202^{+0.034}_{-0.027}$, with
$\chi^2/\text{d.o.f.}\approx 1.09$.  Thus the fit remains good
and the value obtained for $\Lambda_{\overline{MS}}^{\left(5\right)}$ is
stable under variation of $\mu_I$, while the value of $\alpha_0$
decreases as expected for a running effective coupling.  Indeed,
the implied mean value of $\alpha_{\text{eff}}$ in the range 2-3 GeV,
\begin{equation}
\overline\alpha_{\text{eff}} = 3\,\alpha_0\left(3\text{ GeV}\right)
-2\,\alpha_0\left(2\text{ GeV}\right) = 0.19\pm 0.10
\end{equation}
is consistent with the perturbative value
$\alpha_s\left(2.5\text{ GeV}\right)=0.26$.

\subsection{Final comparison with experimental distributions}
\label{sec:final}
\begin{figure}
\begin{center}
\includegraphics[scale=0.51]{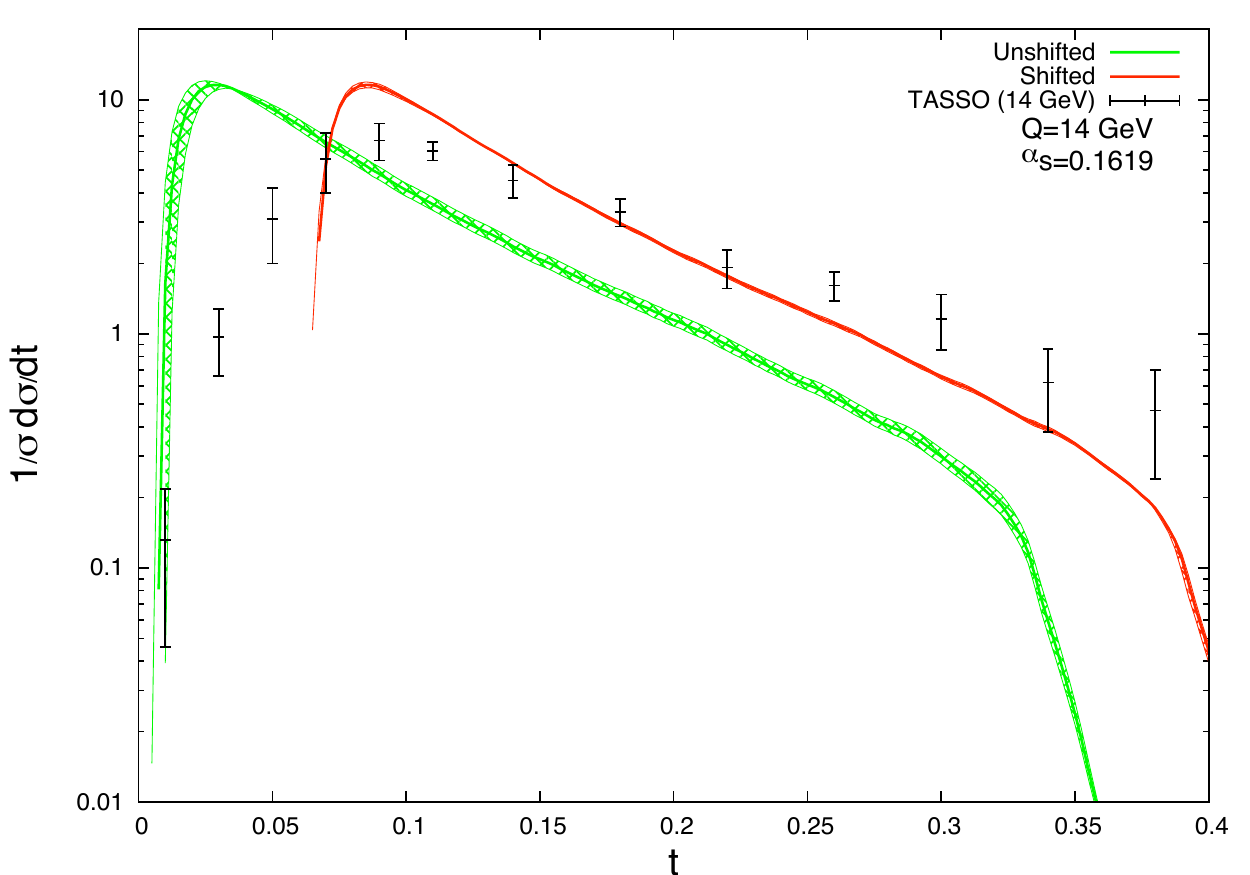}
\includegraphics[scale=0.51]{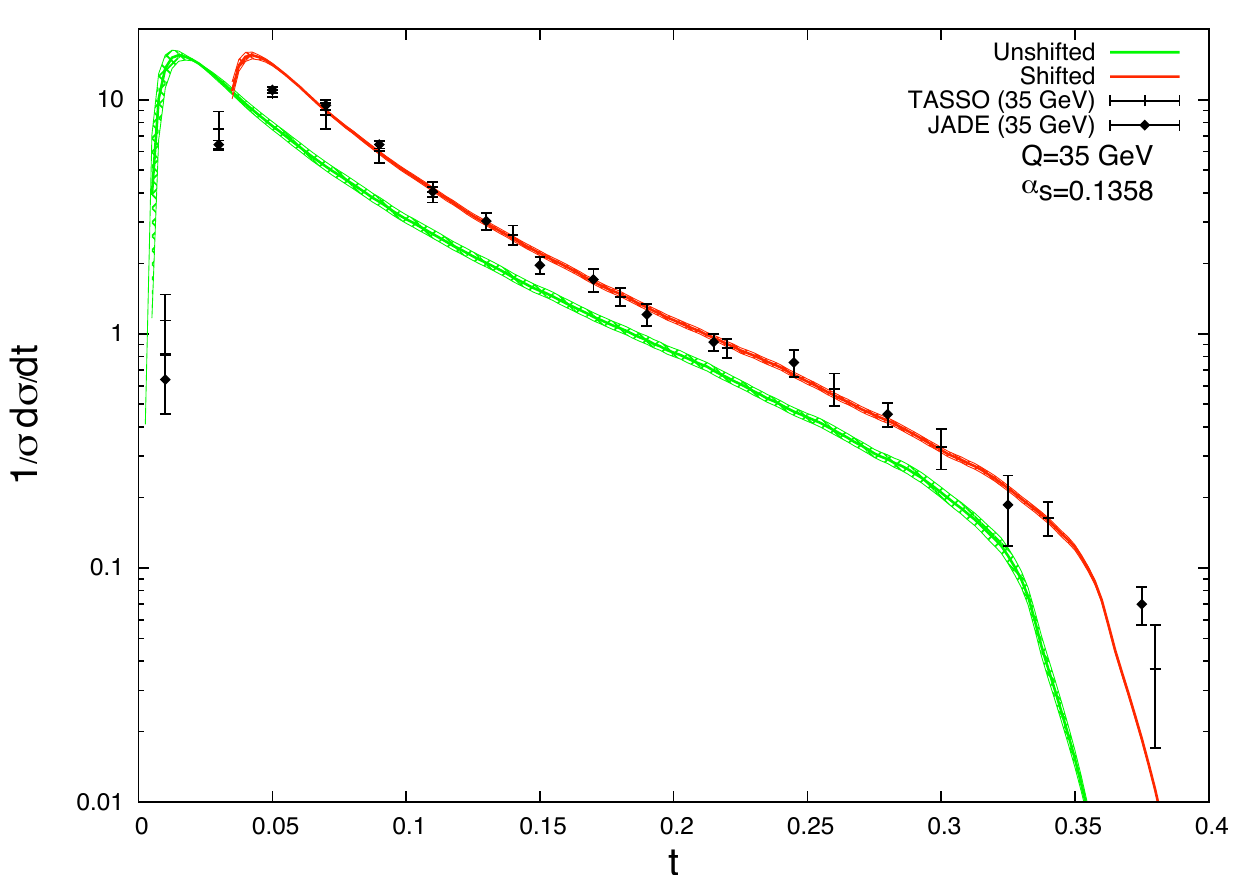}
\includegraphics[scale=0.51]{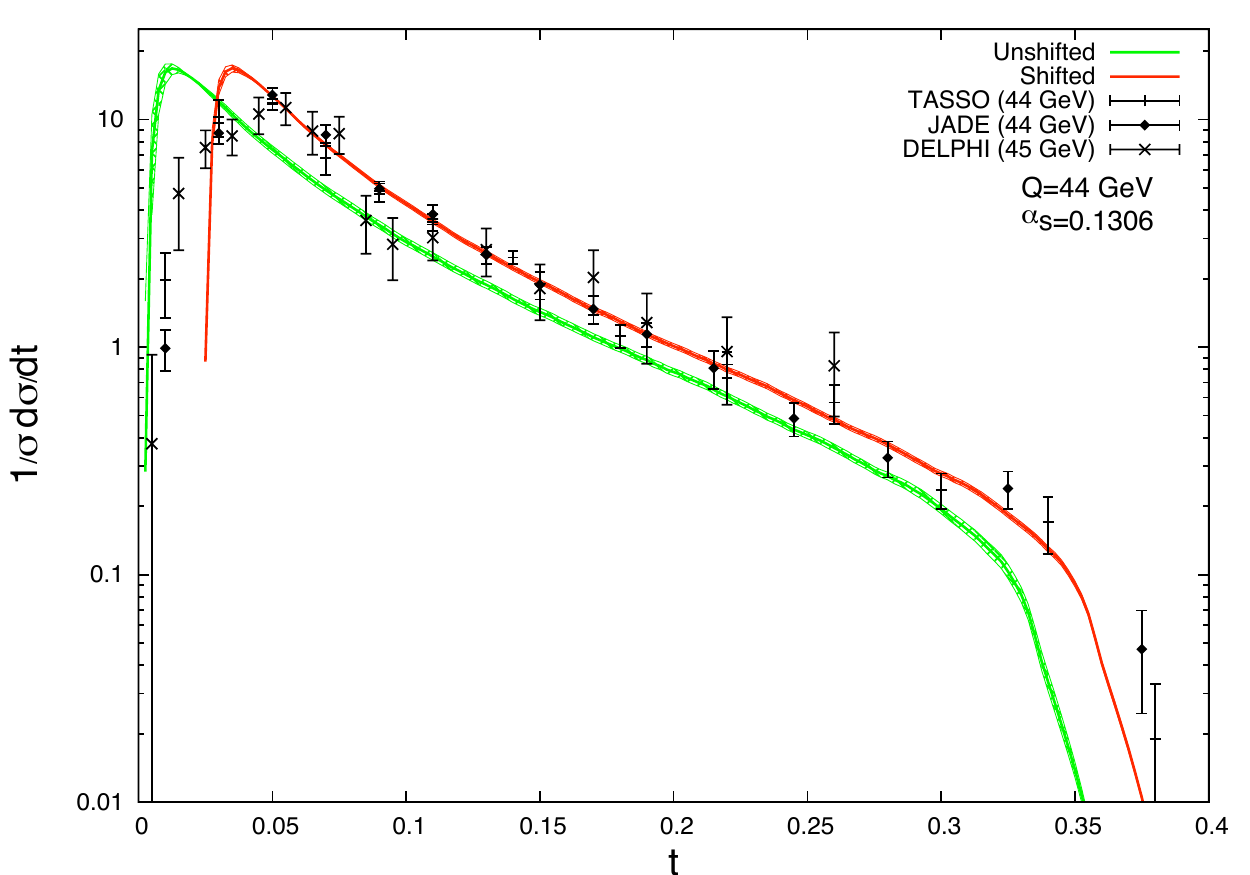}
\includegraphics[scale=0.51]{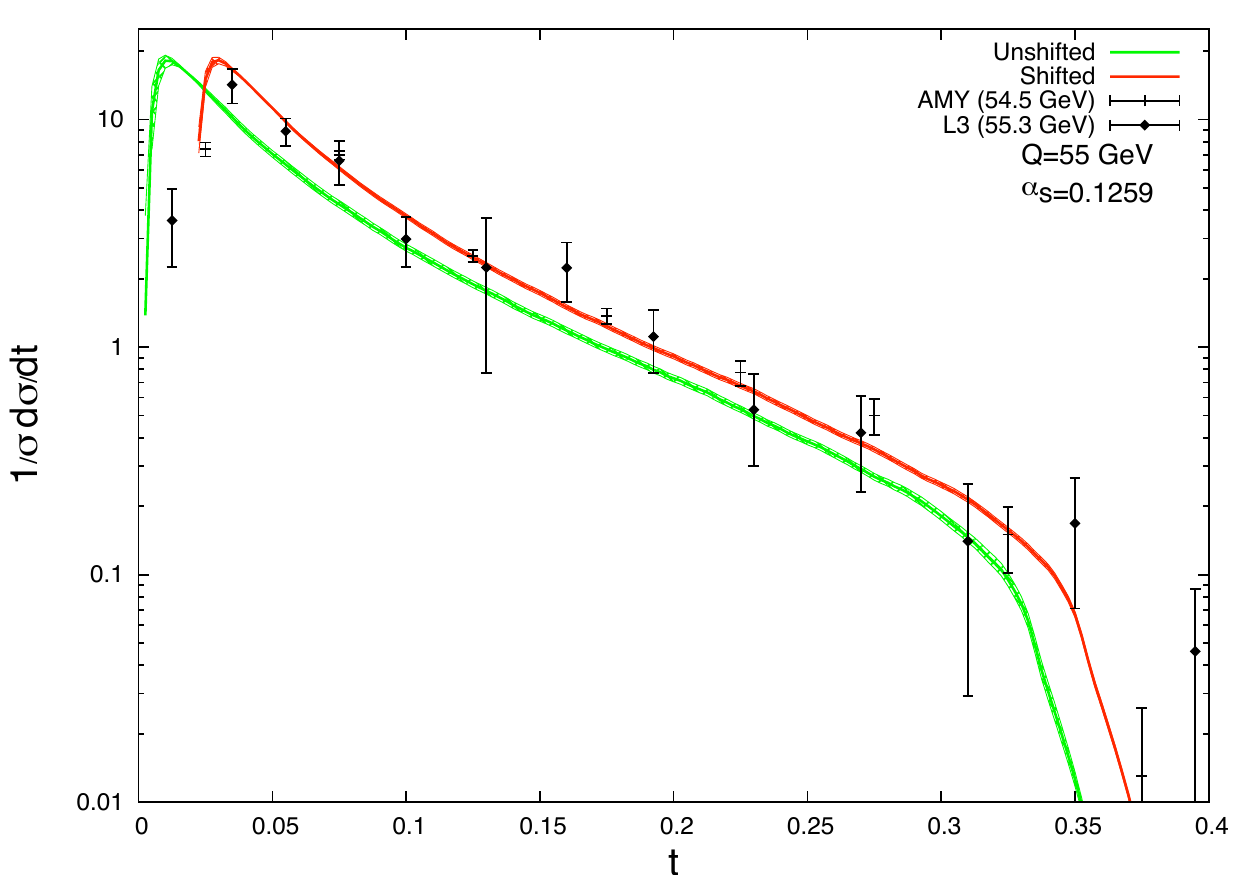}
\includegraphics[scale=0.51]{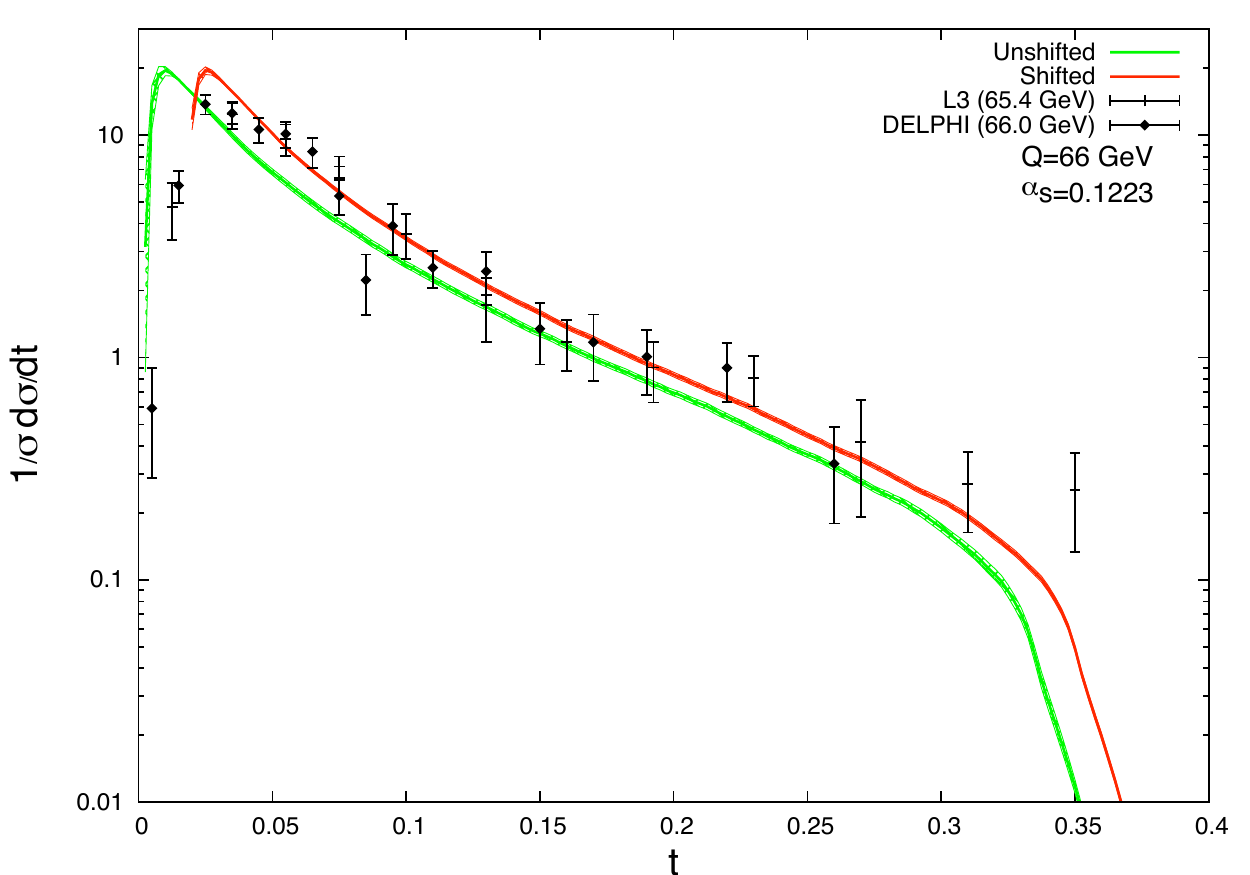}
\caption{Comparison of shifted, unshifted and experimental thrust
  distributions: $Q=14-66$ GeV.\label{fig:shift1}}
\end{center}
\end{figure}

\begin{figure}
\begin{center}
\includegraphics[scale=0.51]{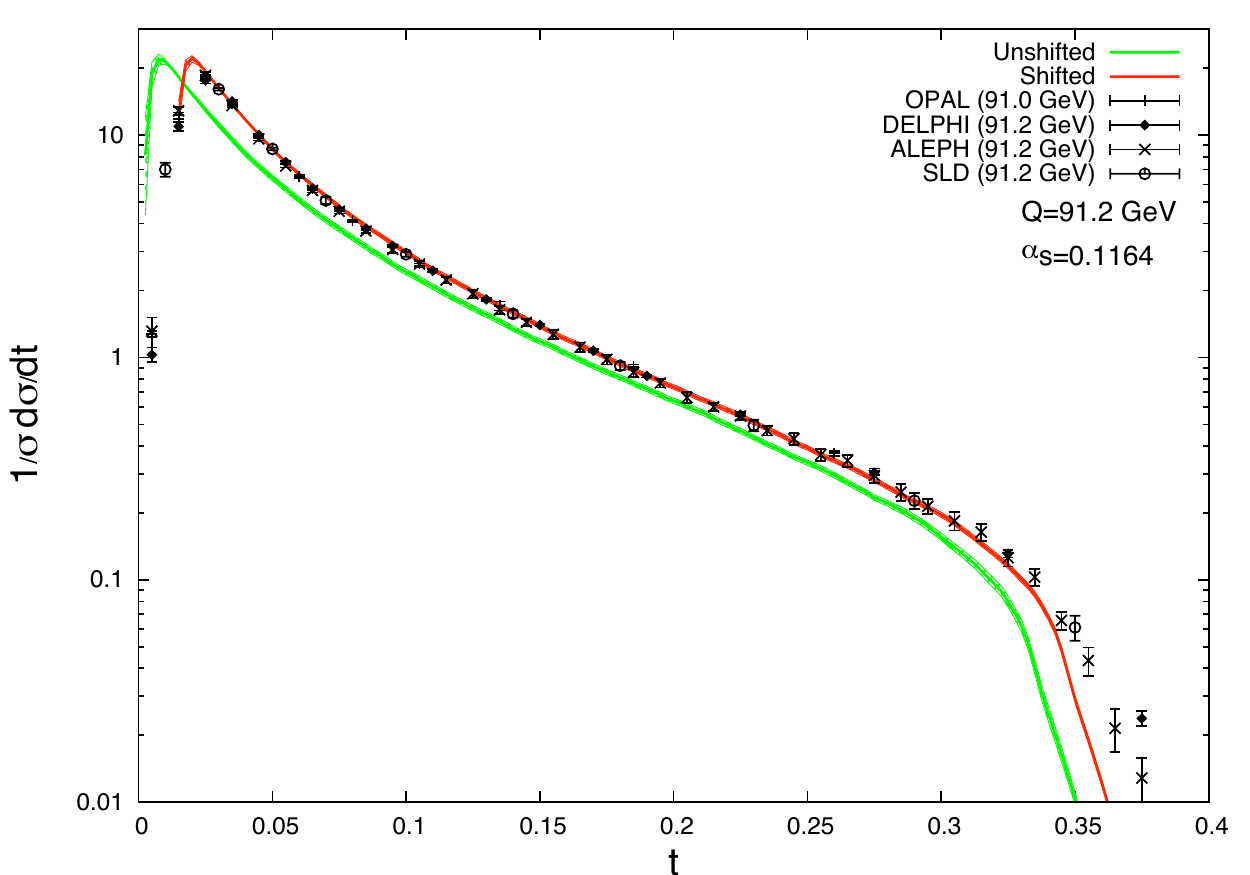}
\includegraphics[scale=0.51]{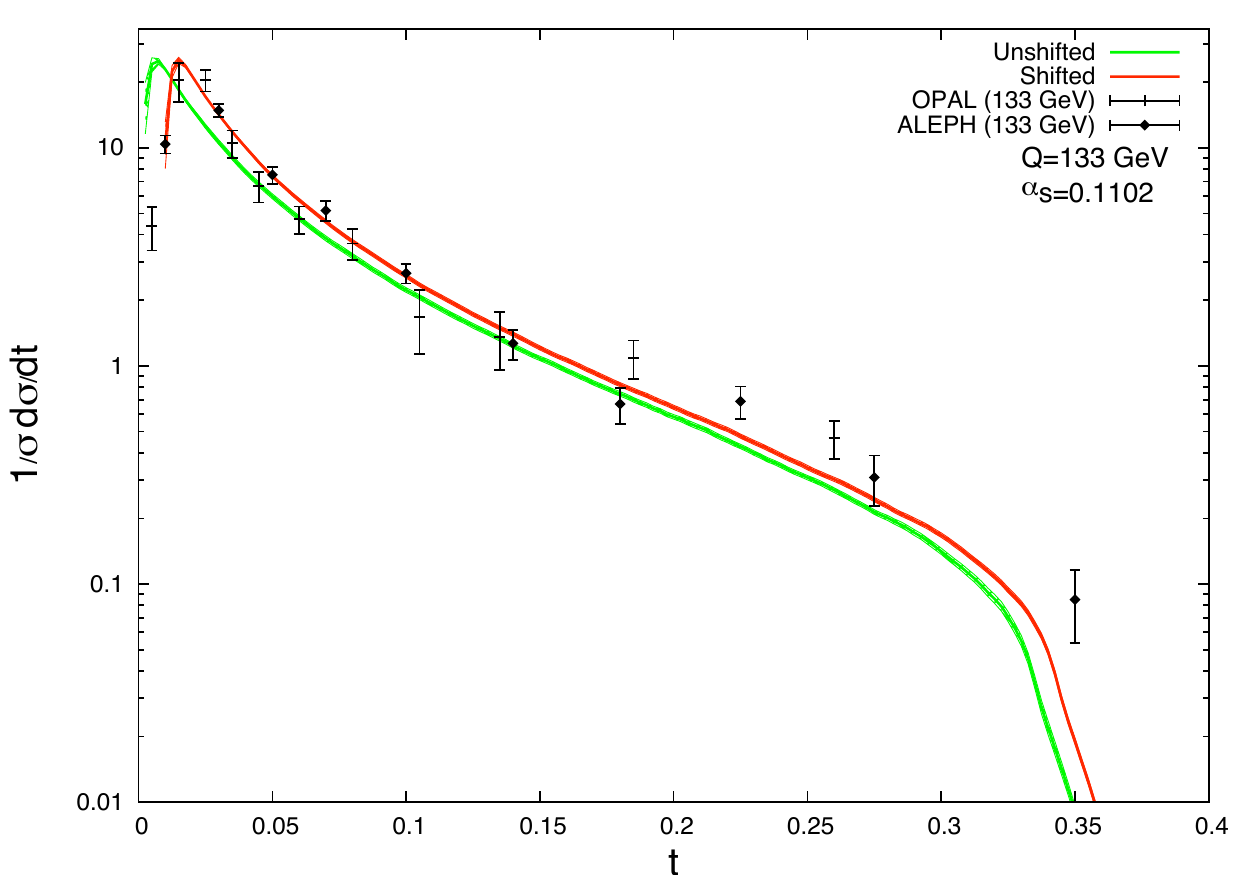}
\includegraphics[scale=0.51]{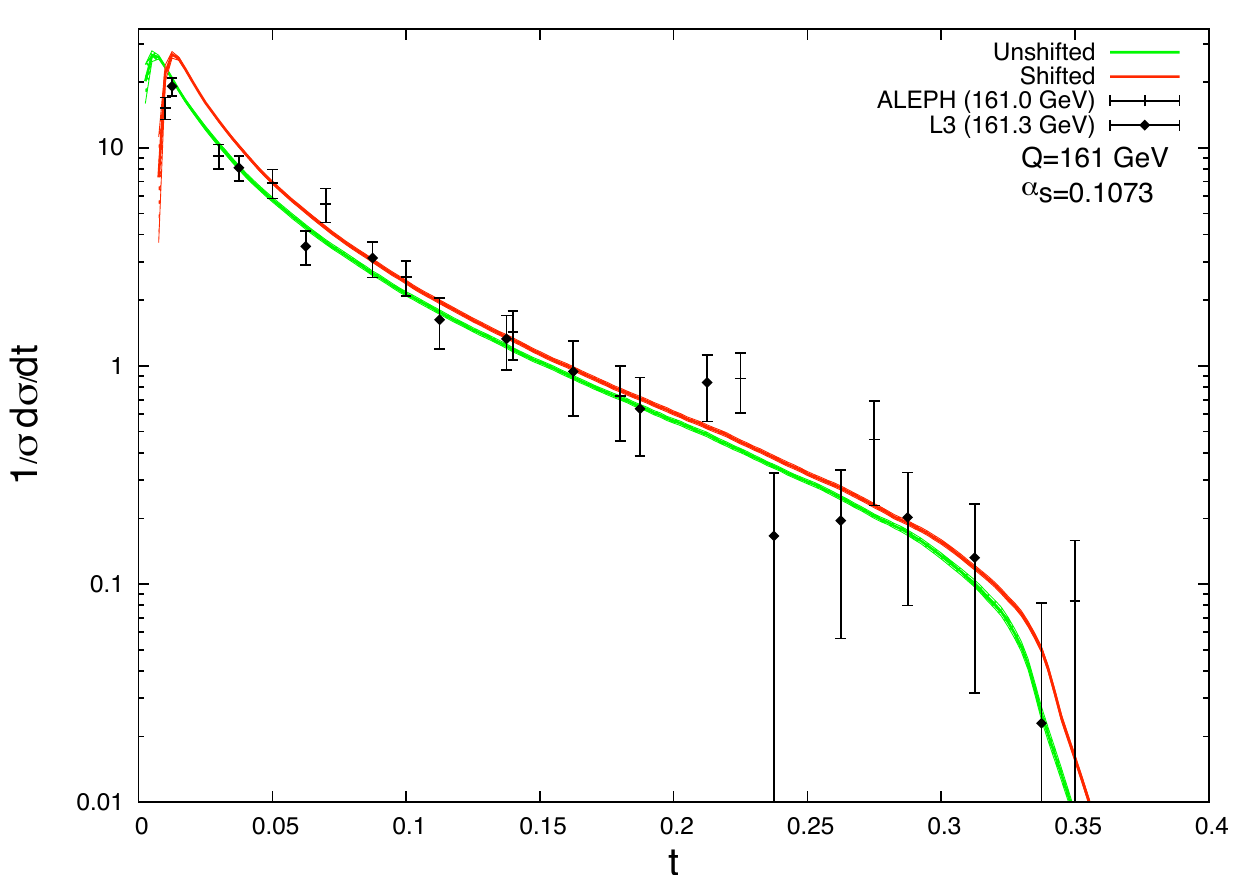}
\includegraphics[scale=0.51]{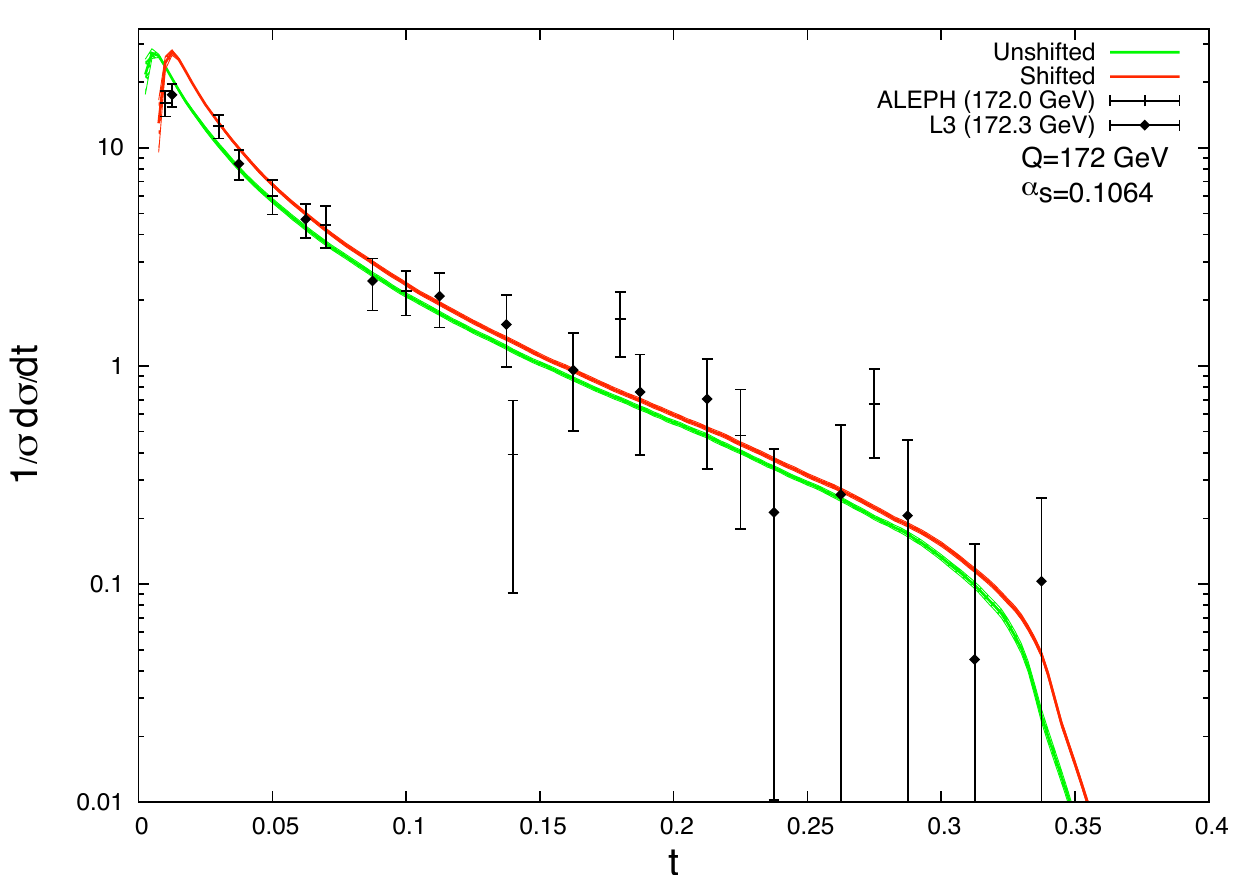}
\includegraphics[scale=0.51]{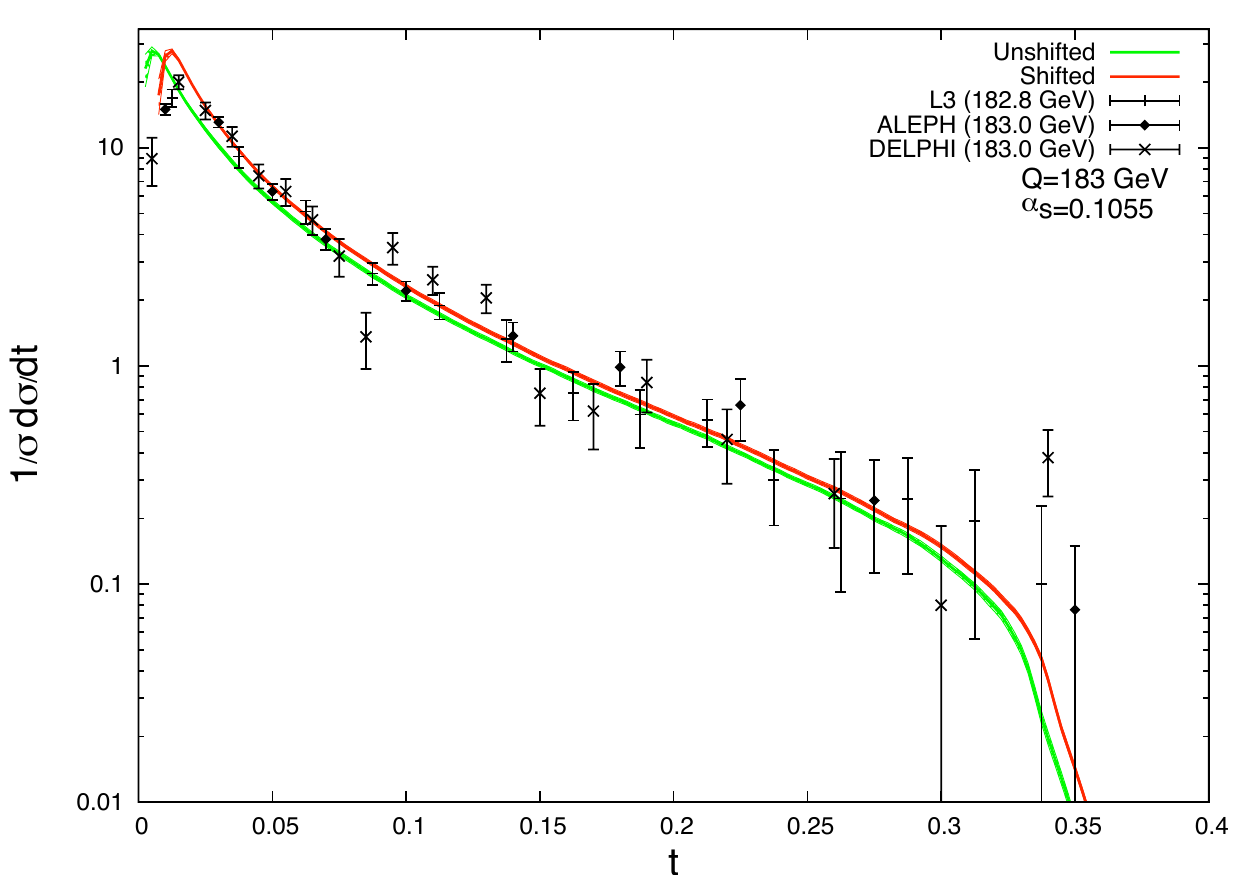}
\caption{Comparison of shifted, unshifted and experimental thrust
  distributions: $Q=91-183$ GeV.\label{fig:shift2}}
\end{center}
\end{figure}

\begin{figure}
\begin{center}
\includegraphics[scale=0.51]{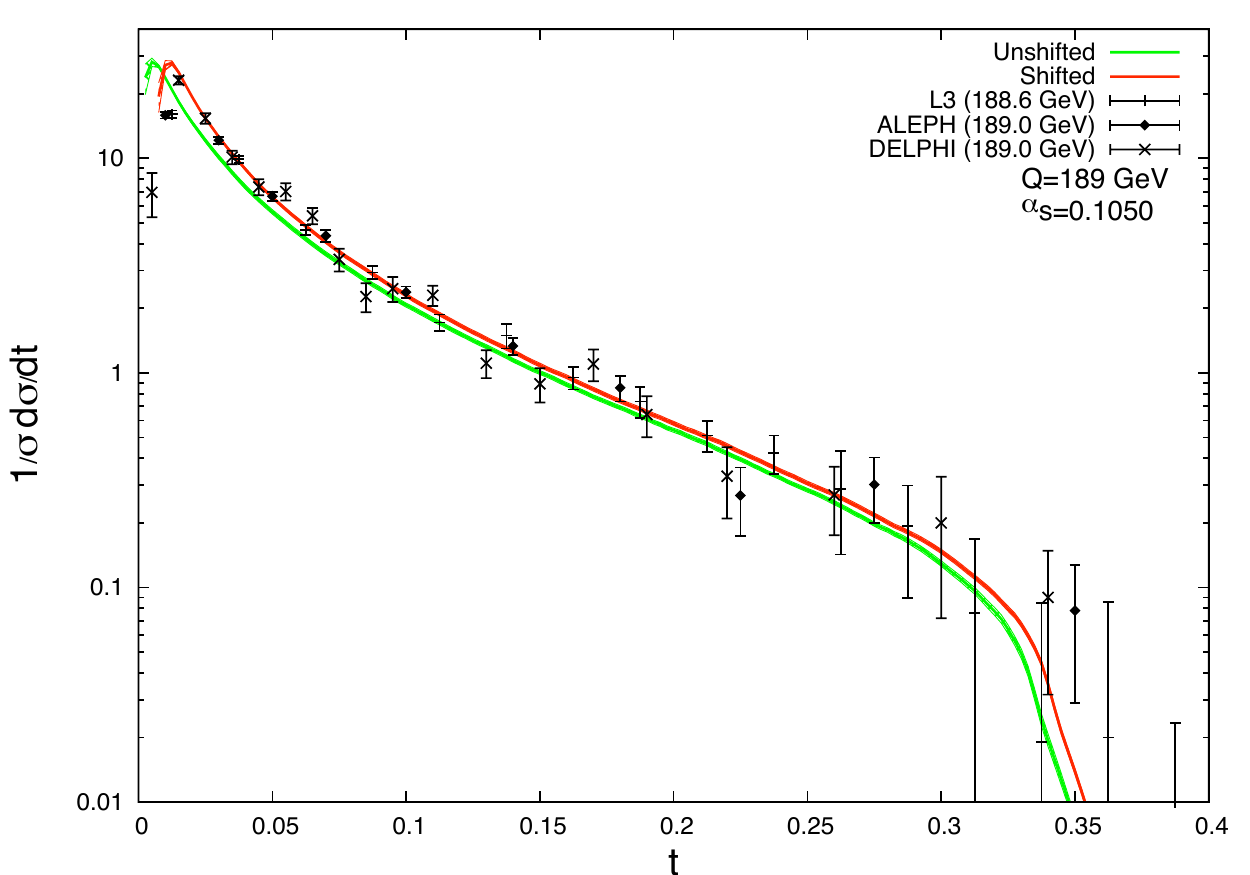}
\includegraphics[scale=0.51]{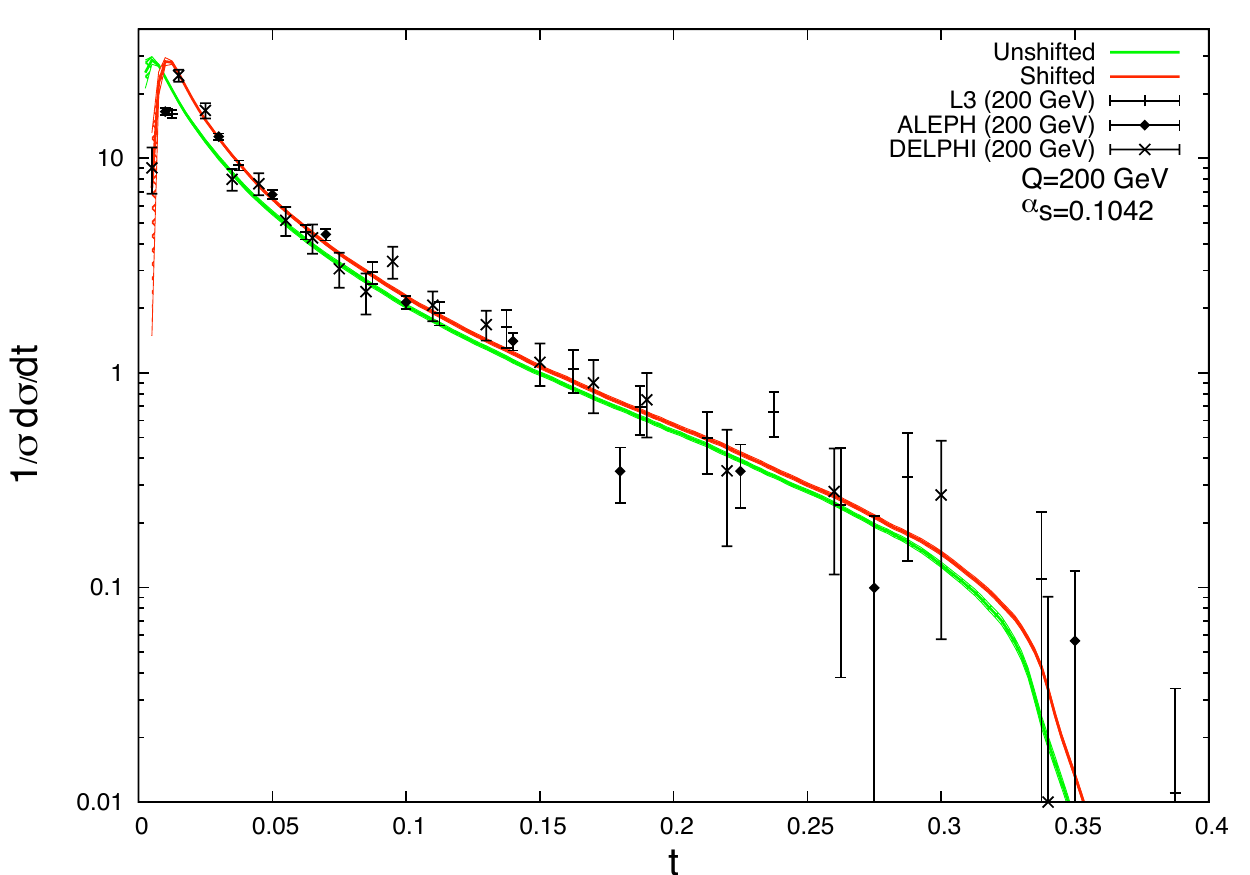}
\includegraphics[scale=0.51]{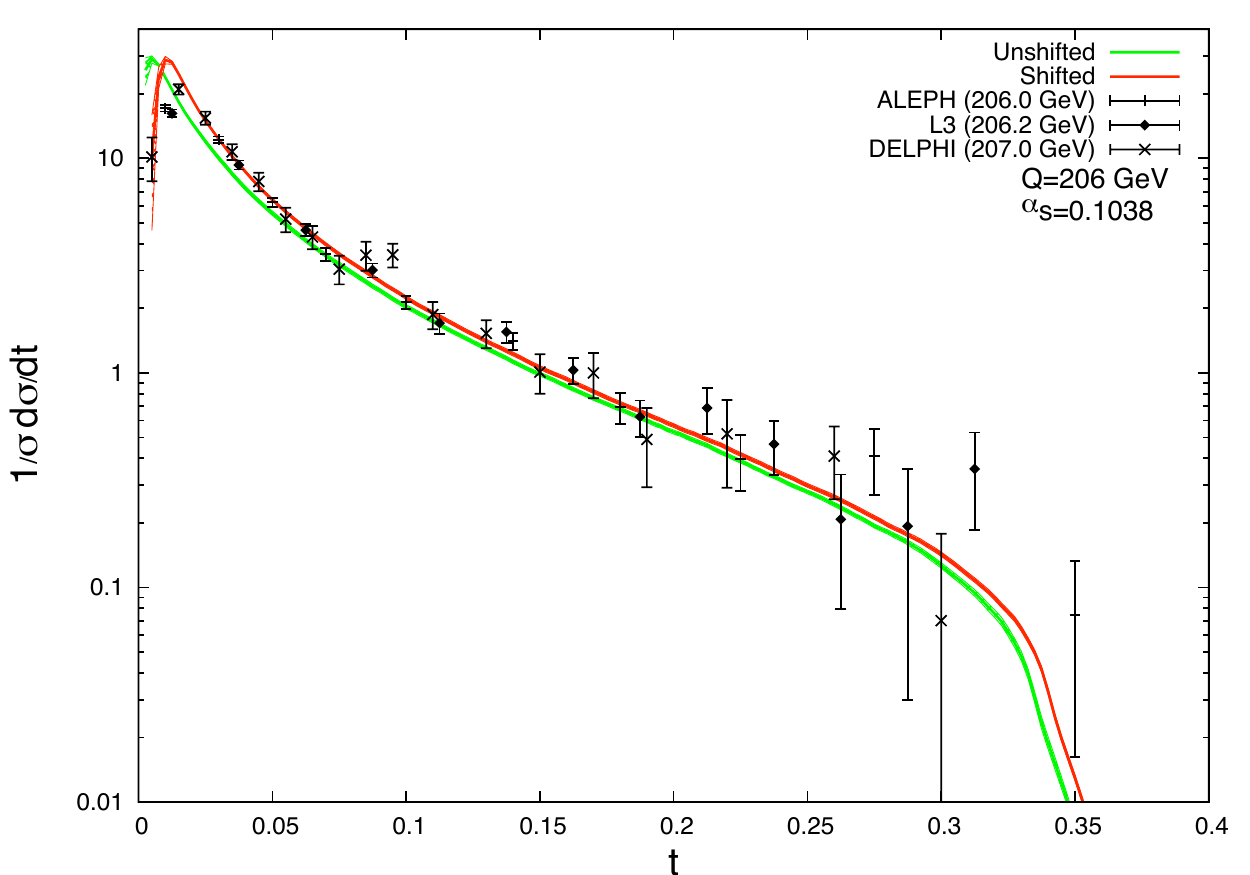}
 \caption{Comparison of shifted, unshifted and experimental thrust
  distributions: $Q=189-207$ GeV.\label{fig:shift3}}
\end{center}
\end{figure}

Figures \ref{fig:shift1}-\ref{fig:shift3} show the final
(NNLO+NLL+shift) theoretical distributions in comparison to the
experimental ones, with the best-fit values of $\alpha_0$ and
$\alpha_s$ assumed. The shaded area around the unshifted distribution
is the renormalisation scale uncertainty found by varying
$\mu_{R}^{2}\in\left[Q^2/2,2Q^2\right]$, and the shaded area around the shifted
distribution is the corresponding error found by varying between the
best fit limits obtained previously ($\alpha_0\left(2\text{
    GeV}\right)=0.585$, $\Lambda_{\overline{MS}}^{\left(5\right)}=0.173$ GeV and
$\alpha_0\left(2\text{ GeV}\right)=0.598$,
$\Lambda_{\overline{MS}}^{\left(5\right)}=0.210$ GeV).


It is clearly seen that inclusion of the shift results in a
significantly more accurate distribution over the fit range,
particularly for the lower energies. As the best fit value of
$\alpha_s$ is very close to the world average, the unshifted
distributions here are essentially the same as those in
Figs.~\ref{fig:unshift1}-\ref{fig:unshift3}.

\section{Conclusions}
\label{sec:conc}
We have seen that the extension of the NNLO perturbative distribution
to NNLO+NLL accuracy results in an improved matching with experiment,
particularly in the low $t$ region.

Analysis of the difference between the perturbative and experimental
distributions over a range of energies showed that $1/Q$ power
corrections were required to account for this difference. Replacement
of the perturbative strong coupling with an effective coupling
below an infra-red matching scale was used to include such
non-perturbative corrections in our theoretical calculation and
resulted in a $1/Q$-dependent shift in the distribution. With best-fit
values $\alpha_0\left(2\text{ GeV}\right)=0.59\pm 0.03$ and
$\alpha_s\left(91.2\text{ GeV}\right)=0.1164^{+0.0028}_{-0.0026}$, this gave a
significantly improved matching with the experimental distributions in
the range $14\text{ GeV}\le Q\le 207\text{ GeV}$. These values are
consistent with those achieved in similar analyses to NLO, as
well as with the world-average value of $\alpha_s$.

The agreement of the $\alpha_0$ and $\alpha_s$ values from the
analysis at NNLO+NLL with those obtained at NLO+NLL is a non-trivial
test of the low-scale effective coupling hypothesis.
The presence of the ${\cal O}(\alpha_s^3)$ term in
Eq.~(\ref{eqsixten}), which amounts to about 80\% of the
${\cal O}(\alpha_s^2)$ term, means that we are not simply adding a $1/Q$
correction to the perturbative result, but rather that we are regularizing
the divergent renormalon contribution by modifying the strong coupling
at low scales. This implies that the explicit non-perturbative $1/Q$
shift applied to the perturbative prediction becomes smaller as higher
orders are computed, and would eventually change sign at sufficiently
high orders, as the renormalon contribution grows indefinitely.

A similar analysis to that in this work could be repeated for other
event shape variables whose distributions have been determined
perturbatively to NNLO and for which resummation of large logarithms
is possible. Perturbative resummed calculations of such distributions
have been performed \cite{twelve} but non-perturbative effects
have not been included in the way advocated here -- they are not
necessarily simple shifts as in the case of thrust.  It would also
be of interest to combine the present approach to non-perturbative
effects with soft-collinear effective theory, which permits the
resummation of next-to-next-to-leading logarithms~\cite{Becher:2008cf}.

\section*{Acknowledgements}
We are grateful to the authors of Refs.~\cite{two,three} for providing
results of their calculations and for helpful comments.
BRW thanks the CERN Theory Group for
hospitality while part of this work was performed.  This research was
supported in part by the UK Science and Technology Facilities Council.


\begin{thebibliography}{8}
\bibitem{one}Y.~L.~Dokshitzer and B.~R.~Webber, Phys.\ Lett.\ B \textbf{404} (1997) 321 [arXiv:hep-ph/9704298].
\bibitem{Beneke:1998ui}
  M.~Beneke,
  Phys.\ Rept.\  {\bf 317} (1999) 1
  [arXiv:hep-ph/9807443].
\bibitem{Beneke:2000kc}
  M.~Beneke and V.~M.~Braun,
  arXiv:hep-ph/0010208.
\bibitem{Dokshitzer:1995qm}
  Y.~L.~Dokshitzer, G.~Marchesini and B.~R.~Webber,
  Nucl.\ Phys.\  B {\bf 469} (1996) 93
  [arXiv:hep-ph/9512336].
\bibitem{two}A.~Gehrmann-De Ridder, T.~Gehrmann, E.~W.~N.~Glover and G.~Heinrich, Phys. Rev. Lett. \textbf{99} (2007) 132002 [arXiv:0707.1285 [hep-ph]].
\bibitem{three}A.~Gehrmann-De Ridder, T.~Gehrmann, E.~W.~N.~Glover and
  G.~Heinrich, JHEP \textbf{0712} (2007) 094 [arXiv:0711.4711 [hep-ph]].
\bibitem{GehrmannDeRidder:2007jk}
  A.~Gehrmann-De Ridder, T.~Gehrmann, E.~W.~N.~Glover and G.~Heinrich,
  JHEP {\bf 0711} (2007) 058
  [arXiv:0710.0346 [hep-ph]].
\bibitem{Weinzierl:2008iv}
  S.~Weinzierl,
  arXiv:0807.3241 [hep-ph].
\bibitem{Amsler:2008zz}
  C.~Amsler {\it et al.}  [Particle Data Group],
  Phys.\ Lett.\  B {\bf 667} (2008) 1.
\bibitem{four}S.~Catani, L.~Trentadue, G.~Turnock and B.~R.~Webber, Nucl.\ Phys.\ B \textbf{407} (1993) 03.
\bibitem{Dokshitzer:1982xr}
  Y.~L.~Dokshitzer, V.~S.~Fadin and V.~A.~Khoze,
  Z.\ Phys.\  C {\bf 15} (1982) 325;
  Z.\ Phys.\  C {\bf 18} (1983) 37.
\bibitem{Bassetto:1984ik}
  A.~Bassetto, M.~Ciafaloni and G.~Marchesini,
  Phys.\ Rept.\  {\bf 100}, 201 (1983).
\bibitem{Catani:1990rr}
  S.~Catani, B.~R.~Webber and G.~Marchesini,
  Nucl.\ Phys.\  B {\bf 349} (1991) 635.
\bibitem{tasso}TASSO Collaboration (W.~Braunschweig \textit{et al.}), Z.\ Phys.\ C \textbf{47} (1990) 187.
\bibitem{jade}JADE Collaboration (P~.A.~Movilla Fernandez \textit{et al.}), Eur.\ Phys.\ J.\ C \textbf{1} (1998) 461 [arXiv:hep-ex/9708034].
\bibitem{l3}L3 Collaboration (P.~Achard \textit{et al.}), Phys.\ Rept.\ \textbf{399} (2004) 71 [arXiv:hep-ex/0406049].
\bibitem{delphi}DELPHI Collaboration (J.~Abdallah \textit{et al.}), Eur.\ Phys.\ J.\ C \textbf{29} (2003) 285 [arXiv:hep-ex/0307048].
\bibitem{amy}AMY Collaboration (Y.~K.~Li \textit{et al.}), Phys.\ Rev.\ D \textbf{41} (1990) 2675.
\bibitem{opal}OPAL Collaboration (G.~Abbiendi \textit{et al.}), Eur.\ Phys.\ J.\ C \textbf{40} (2005) 287 [arXiv:hep-ex/0503051].
\bibitem{aleph}ALEPH Collaboration (A.~Heister \textit{et al.}), Eur.\ Phys.\ J.\ C \textbf{35} (2004) 457.
\bibitem{sld}SLD Collaboration (K.~Abe \textit{et al.}), Phys.\ Rev.\ D \textbf{51} (1995) 962 [arXiv:hep-ex/9501003].
\bibitem{Solovtsov:1997at}
  I.~L.~Solovtsov and D.~V.~Shirkov,
  Phys.\ Lett.\  B {\bf 442} (1998) 344
  [arXiv:hep-ph/9711251];
  Theor.\ Math.\ Phys.\  {\bf 150} (2007) 132
  [arXiv:hep-ph/0611229].
\bibitem{nine}B.~R.~Webber, JHEP \textbf{9810} (1998) 012 [arXiv:hep-ph/9805484].
\bibitem{MovillaFernandez:2001ed}
  P.~A.~Movilla Fernandez, S.~Bethke, O.~Biebel and S.~Kluth,
  Eur.\ Phys.\ J.\  C {\bf 22} (2001) 1
  [arXiv:hep-ex/0105059].
\bibitem{eleven}G.~Dissertori, A.~Gehrmann-De Ridder, T.~Gehrmann, E.~W.~N.~Glover, G.~Heinrich and H.~Stenzel, JHEP \textbf{0802} (2008) 040 [arXiv:0712.0327 [hep-ph]].
\bibitem{twelve}T.~Gehrmann, G.~Luisoni and H.~Stenzel, Phys.\ Lett.\ B \textbf{664} (2008) 265 [arXiv:0803.0695 [hep-ph]].
\bibitem{Becher:2008cf}
  T.~Becher and M.~D.~Schwartz,
  JHEP {\bf 0807} (2008) 034
  [arXiv:0803.0342 [hep-ph]].
\end{thebibliography}
\end{document}